\documentclass{aa}
\usepackage[varg]{txfonts}
\usepackage{graphicx}
\usepackage{natbib}
\usepackage{grffile} % for files with underscores
\usepackage{epstopdf}
\usepackage{color}
\usepackage{ulem}

\definecolor{blue}{rgb}{0,0.0,0.9}

\begin{document}

\title{Atmospheric effects of stellar cosmic rays on Earth-like exoplanets orbiting M-dwarfs}
\author{F. Tabataba-Vakili \inst{\ref{inst1},\ref{inst2}}
\and J. L. Grenfell \inst{\ref{inst3}}
\and J.-M. Grie\ss meier \inst{\ref{inst4},\ref{inst5}}
\and H. Rauer\inst{\ref{inst1},\ref{inst3}}}

\institute{Zentrum f\"ur Astronomie und Astrophysik, Technische Universit\"at Berlin, Hardenbergstraße 36, 10623 Berlin, Germany \label{inst1}
\and
Now at: Atmospheric, Oceanic and Planetary Physics, Department of Physics, University of Oxford, Oxford OX1 3PU, UK \email{Fachreddin.Tabataba-Vakili@physics.ox.ac.uk}\label{inst2}
\and Institut f\"ur Planetenforschung, Deutsches Zentrum f\"ur Luft- und Raumfahrt, Rutherfordstraße 2, 12489 Berlin, Germany \label{inst3}
\and Laboratoire de Physique et Chemie de l'Environnement et de l'Espace, Universit\'e d'Orl\'eans / Centre National d'Etudes Spatiales, France \label{inst4}
\and Station de Radioastronomie de Nancay, Observatoire de Paris - Centre National d'Etudes Spatiales/Institut national des sciences de l'Univers, USR 704 - Universit\'e Orl\'eans, Observatoire des Sciences de l'Univers en région Centre, route de Souesmes, 18330 Nancay, France \label{inst5}}

\abstract{M-dwarf stars are generally considered favourable for rocky planet detection.
However, such planets may be subject to extreme conditions due to
possible high stellar activity. 
The goal of this work is to determine the potential effect of stellar
cosmic rays on key atmospheric species of Earth-like planets
orbiting in the habitable zone of M-dwarf stars and show corresponding changes
in the planetary spectra. 
We build upon the cosmic rays model scheme of %\citet{grenfell2012response}
Grenfell et al. (2012), who considered cosmic ray induced NOx production, by
adding further cosmic ray induced production mechanisms (e.g. for HOx) and
introducing primary protons of a wider energy range (16 MeV - 0.5
TeV).
Previous studies suggested that planets in the habitable zone that are subject to strong flaring conditions have 
high atmospheric methane concentrations, while their ozone biosignature is completely destroyed.
Our current study shows, however, that adding 
cosmic ray induced HOx production can cause a decrease in atmospheric
methane abundance of up to 80\%.  Furthermore, the cosmic ray induced  HOx molecules react with NOx to produce HNO$_3$, which 
produces strong  HNO$_3$ signals in the theoretical spectra and reduces NOx-induced catalytic destruction of ozone  so that more than 25\% of the ozone column remains.
Hence, an ozone signal  remains visible in the theoretical spectrum (albeit with a weaker intensity) when
incorporating the new cosmic ray induced NOx and HOx schemes, even for a
constantly 
flaring M-star case. We also find that HNO$_3$ levels may be high enough to be potentially detectable. Since ozone concentrations, which act as the key shield against harmful UV radiation, are affected by cosmic rays via NOx-induced catalytic destruction of ozone,  the impact of stellar cosmic rays on surface UV fluxes is also studied.
}


\authorrunning{Tabataba-Vakili et al.}
\titlerunning{Atmospheric effects of stellar cosmic rays on exoplanets orbiting M-dwarfs}

\maketitle

\section{Introduction}

With the discovery of the first exoplanet around a main-sequence star \citep{mayor1995jupiter}, the scope of the search for life has extended beyond our solar system. 
As of today, nearly 2000 exoplanets  have been detected, over 1200  of  these  via the transit method, in which the occultation of the star by the planet can be measured photometrically.\footnote{see \url{http://exoplanet.eu/}} 
Transiting exoplanets are of interest for the detection of life, since it is theoretically possible to characterise their atmosphere spectroscopically for sufficiently bright targets. 
This would allow the detection of chemical compounds that are associated with life forms, termed biosignatures.

Water plays a crucial role in the development and sustenance of life as we know it. This has led to the concept of the habitable zone  \citep[HZ; see][]{huang1959occurrence,kasting1993habitable}. The HZ is defined as the orbital distance range of the planet where its surface may maintain water in liquid form. This range depends e.g. on the planet's atmospheric composition and the temperature of its host star. Herein lies one advantage of the comparatively cool M-dwarf host stars: habitable planets are situated closer to their host, which improves the geometric probability of a transit. Other favourable characteristics of M-dwarf hosts include their small size, which improves detectability of Earth-sized planets in the HZ, and the large abundance of M-dwarf stars in our Galactic vicinity \citep{scalo2007m}. 

%talk more about CR work in the exoplanet and Earth literature (Lee, Segura 2010,  Jackman etc.). Why it is interesting, what are main effects, what is lacking in the literature!! 
%Cr effect on life

Since a habitable planet orbiting an M-dwarf star lies so close to its host and because the planet might have a weak magnetosphere (due to tidal-locking), the planet may be subject to high particle fluxes \citep{griessmeier2005cosmic}, which are termed cosmic rays (CRs). 
These CRs are categorised by origin: Galactic cosmic rays (GCRs) originate outside the Solar System, while solar (or stellar) cosmic rays (SCRs) originate from the Sun (or the host star of the studied planet). 
Independent of their origin, CRs can interact with the planetary atmosphere, possibly destroying important biosignature species or affecting the radiative balance of the atmosphere by destroying key radiative species. % If strong enough, cosmic rays could even strip planets of their atmosphere due to loss processes (Lammer).

For Earth, extensive research has gone into understanding the ionizing effects of SCRs and solar proton events on the chemistry of middle atmosphere \citep[e.g.][]{jackman1990effect,jackman2005neutral,winkler2008modeling}. In this region CRs are one of the key chemical forcing mechanisms. On the other hand, GCRs provide a constant forcing on the atmospheric chemistry over all altitudes. \citet{nicolet1975production} and \citet{solomon1981effect} have studied the rate of CR-induced NOx and HOx production. Recently, \citet{calisto2011influence} have analysed the effect of GCRs on the global atmospheric chemistry with a 3D chemistry-climate model. They suggest that on Earth GCRs can locally produce up to 30\% of available NO, causing a loss of up to 3\% in the ozone column.
%\textbf{[Literature Review]}

%Due to the favorable detection criteria \citep{scalo2007m}, the habitability of M-dwarf planets has been extensively researched. Some studies e.g.  \cite{joshi2003climate}  suggest that atmospheres of tidally-locked planets remain dynamically stable (favoring habitability). Other studies e.g. \cite{lammer2011loss} claim that nitrogen-rich atmospheres may be eroded through loss processes during the early stages of planetary evolution (disfavoring habitability).  

Owing to their favourable detection criteria \citep{scalo2007m}, the habitability of M-dwarf planets has been extensively studied. For instance,  
dynamical three-dimensional studies suggest that tidally-locked planets, which are expected to be common in the HZ of M-dwarfs, have regions that may be able to sustain habitable conditions \citep{joshi2003climate}. 
Exoplanet parameter studies using GCMs focus largely on quantifying the effect of certain parameters on the planet's habitability and general circulation. %, mostly in terms of surface temperatures, 
These vary for instance the orbital eccentricity \citep{dressing2010}, the planetary obliquity \citep{spiegel2009}, the surface roughness \citep{rauscher2012}, or the stellar spectrum \citep{godolt2012}.
Further recent GCM studies, e.g. by \citet{yang2013stabilizing}, propose that clouds on tidally-locked exoplanets provide a stabilizing feedback, expanding the M-dwarf habitable zone. 
%Other recent parameter studies that use simplified GCMs aim to define a dimensionless parameter space, and define key parameters that sufficiently characterise the general circulation of their idealized atmospheric models. Such an approach utilizes the similarity principle of fluid-dynamic systems to identify circulation regimes of exoplanets in a more general manner. In specific, studies such as these focus on identifying e.g. the wave-forcing responsible for the emergence of equatorial superrotation \citep{potter2014}, the effect of a parameterised seasonal heating on superrotation \citep{mitchell2014}, or focussing on thermodynamic diagnostics \cite{pascale2013}, characterizing jet formaltion mechanisms \citep{wang2014}, as well as setting up a regime diagram of expected circulation patterns in certain points of dimensionless parameter space.
Studies of atmospheric loss processes \citep[e.g.][]{lammer2011loss} suggest that planets with nitrogen-rich atmospheres orbiting M-dwarfs may be eroded during the early stages of planetary evolution unless the planet was protected by a strong intrinsic magnetic field.
%\comm{I think we should add ``unless the planet was protected by a strong intrinsic magnetic field.'' Please check.}

%3D studies
%\cite{joshi2003climate} studied the dynamcial stability of tidally-locked 

%Some studies favor habitability e.g.  \cite{joshi2003climate}  conclude that atmospheres of tidally-locked planets remain dynamically stable; while others disfavor habitability e.g. \cite{lammer2011loss}, who suggest that nitrogen-rich atmospheres may be eroded through loss processes during the early stages of planetary evolution.

%\textbf{[MORE!]}

%chemistry studies

%detectability studies

The detectability of biosignatures in M-dwarf planet atmospheres is directly linked to the planet's atmospheric chemistry.
Key studies that address biosignature detectability   \citep[e.g.][]{selsis2000,selsis2002,des2002,segura2003ozone,segura2005biosignatures,tinetti2006,ehrenreich2006,kaltenegger2007,kaltenegger2009,kaltenegger2010,rauer2011potential,tessenyi2013}  investigate the impact of varying e.g. the stellar spectrum, the abundance of radiative species, or background atmospheres
on synthetic  emission and transmission spectra of hypothetical exoplanets.
%by producing synthetic spectra of hypothetical exoplanets are

With regard to the photochemistry of exoplanets orbiting M-dwarfs,  \citet{segura2005biosignatures} studied the effect of M-class stellar spectra on the atmospheric chemistry, which suggested strongly increased atmospheric concentrations of CH$_4$ and H$_2$O 
with the assumption of an Earth-like atmosphere (i.e. N$_2$, O$_2$ dominated, 1 bar surface pressure). 
 A recent study by \citet{tian2014} suggested that abiotic production of O$_2$ from CO$_2$ photolysis could be significant for planets orbiting M-dwarf stars.
%\textbf{\citet{grenfell2013} }
%\comm{add grenfell2013?, segura 2010?}
%\comm{increased with respect to what?} 
 \citet{grenfell2007biomarker} investigated the effect of GCRs on biomarker molecules for biogenic Earth-like planets surrounding  an active M-dwarf with varying  heliospheric and planetary magnetic shielding. This study was recently built on by \citet{griessmeier2014} and \citet{griessmeier2014a}, who use the updated version of the CR model presented in the current work to examine the effect of the planetary magnetospheric shielding  on GCR-induced photochemistry. With a maximum of 20\% difference in column-integrated ozone values (and 6\% for CH$_4$ and H$_2$O) between no magnetic shielding and ten times the Earth's magnetic field strength, they find GCRs have nearly no effect on the resulting planetary spectra.
A further study by \citet{grenfell2012response} focused on the effect of high-intensity SCR fluxes on the atmospheric chemistry via the mechanism of CR-induced NO production. They concluded that SCRs are able to significantly reduce  the O$_3$ column by 99.99\% for highly flaring stars via NOx-induced catalytic destruction.

This work focuses on the photochemical response to stellar cosmic rays and the corresponding effects upon theoretical planetary spectra and surface UV radiation of an Earth-like planet orbiting an M-dwarf star. 
The main aim of the current study is to improve upon the cosmic ray modelling scheme provided by \citet{grenfell2012response}, thereby expanding the energy range of incident particles and adding further CR-induced photochemical mechanisms (NOx and HOx production). The improved model is then used to provide new insights into the biomarker chemistry of Earth-like exoplanets.  
We then study the effect of changes in the atmospheric chemistry on the spectral signals of key biosignature species, and evaluate surface UV fluxes.

The UV flux on the planetary surface can have harmful consequences to life forms and therefore affects the habitability of a planet. The UV surface flux is strongly affected by the interplay of top of atmosphere (TOA) fluxes and atmospheric O$_3$ concentrations. \citet{segura2010effect} have modelled the photochemical effects of a large UV flare observed on AD Leonis (AD Leo), and find that UV surface fluxes stay within non-lethal dosages. \citet{griessmeier2014} find that GCR-induced O$_3$ destruction results in up to 40\% increased biologically weighted UV surface flux. However, as the AD Leo spectrum used features lower UV-A and UV-B fluxes and the planetary atmosphere retains enough O$_3$ to shield from UV-C, total UV fluxes are much lower than on Earth. 
When using the flaring UV spectrum of \citet{segura2010effect} for both ``long'' and ``short'' flare scenarios, \citet{griessmeier2014} find that the GCR-induced O$_3$ reduction would not be harmful to life with maximum biologically weighted UV surface fluxes not exceeding four-times Earth values. \citet{griessmeier2014} used a similar model version as this work i.e. without time dependence. ``Long'' flares were assumed to maintain the atmosphere in a continuously perturbed state, whereas for ``short'' flares it was assumed that the atmosphere did not respond photochemically.
\citet{grenfell2012response} suggests that for planets in the HZ of M-dwarfs, the effect of SCRs to atmospheric O$_3$ concentrations is more devastating and can cause nearly full atmospheric transmissivity of UVB radiation. The addition of new CR-induced photochemical mechanisms in the current work has 
a significant effect on O$_3$ concentrations and hence has an effect on UV fluxes. Consequently, we evaluate the effect of SCRs to biologically weighted UV surface fluxes using the  scenarios presented in \citet{griessmeier2014}.

This paper is organised as follows:  Section \ref{sec:modd} describes the models used. The model scenarios are introduced in Section \ref{sec:crs}. In Section \ref{sec:atm} the radiative-convective/photochemical model is described. 
The improvements introduced to the cosmic ray scheme are detailed in Section \ref{sec:imp}.
Results regarding atmospheric changes due to the incident SCRs are presented in Sect.~\ref{sec:ap}. In Sect.~\ref{sec:SPE2} significant chemical mechanisms are identified. The resulting planetary emission and transmission spectra are presented in Sect.~\ref{sec:plaspec}. In Sect.~\ref{sec:sflux} the UV flux at the planet's surface is investigated. Sections \ref{sec:disc} and \ref{sec:summ} provide a discussion and concluding remarks.

\section{Model description} \label{sec:modd}

Analogous to \citet{grenfell2012response}, we first compute the primary cosmic ray proton flux at TOA from measurements of the CR flux of the Sun outside the Earth's magnetic field. For the Earth reference, we use a GCR particle spectrum and we assume an Earth-equivalent magnetic field and shielding  using the magnetospheric model of \citet{griessmeier2005cosmic,griessmeier2009protection}.
For the M-dwarf cases we assume a planet without magnetic field shielding, which gives an upper limit to the effect of the CR flux. 
The SCR flux is scaled quadratically with distance to the HZ of the M-dwarf (see Sect.\:\ref{sec:crs}), and is then input into the TOA of the
photochemical part of our 1D radiative convective photochemistry model (see Sect.\:\ref{sec:atm}), where an air-shower approach using a Gaisser-Hillas scheme \citep[see][]{grenfell2007biomarker} is applied to calculate secondary electron fluxes (see Sect.\:\ref{sec:imp}). Contrary to \citet{grenfell2012response} who only calculate NO molecules from these electron fluxes, we additionally introduce CR-induced NOx (N, NO) and HOx (H, OH) production using parametrisations of the ion pair production rate (IPR) found in \citet{jackman2005neutral} and \citet{solomon1981effect}, respectively (see Sect.\:\ref{sec:ipr}). The converged steady-state solution of the atmospheric model is then inserted into a spectral line-by-line model called SQuIRRL \citep{schreier2001a,schreier2003mirart} to calculate theoretical emission and transmission spectra (see Sect.\:\ref{sec:spec}). 

\subsection{Cosmic ray spectra and model scenarios} \label{sec:crs}

Similar to \citet{grenfell2012response}, we look at a number of scenarios, ranging from a planet without CRs to a planet exposed to a flaring M-dwarf star. 
For the M-star cases, the planet is positioned at a distance of 0.153 AU to its host star such that the total irradiance received by the planet is equal to one solar constant. For the stellar spectrum of the host we use the well-studied spectrum of  AD Leonis (AD Leo), which is a strongly flaring M-type star, to provide comparison to previous works. A discussion of the effects of different host spectra can be found in Sect.~\ref{sec:disc}.

Little is known about the SCR fluxes of M-dwarf stars. \citet{khodachenko2012} (their section 2) summarise current uncertainties.
Most studies that investigate the effects of SCRs upon Earth-like atmospheres therefore use solar values for the stellar wind parameters
\citep[e.g.][]{grenfell2012response,segura2010effect}
More recently, 3D Magnetohydrodynamic (MHD) models have been applied to estimate stellar winds on M-dwarfs and their interaction with
the planetary magnetosphere \citep[see e.g.][]{vidotto2013,vidotto2015}, which suggest a strong particle output
for active M-dwarf stars.

\begin{enumerate} 
\setcounter{enumi}{-1}
\item \textbf{Earth reference case}: For the reference case, we use the Galactic cosmic ray reference spectrum outside the magnetosphere given by \citet{seo1994study}. The cosmic ray flux at the top of the atmosphere is calculated by numerically simulating the transport of these particles through an Earth equivalent magnetosphere using the magnetospheric model described in  \citet{griessmeier2005cosmic,grenfell2007biomarker,griessmeier2014a}. 
\label{enum:earth}
\item \textbf{M-dwarf w/o CR}: In this case, CRs are absent.  \label{enum:ADL}
\item \textbf{quiescent M-dwarf}: This case is based on the SCR proton spectrum measured at solar minimum, using the cosmic ray spectrum of \citet{kuznetsov2005models} at 1 AU (their Fig.~2, solid line). No shielding by a planetary magnetic field is assumed. \label{enum:SCR}
\item \textbf{active M-dwarf}: This case is based on the SCR proton spectrum measured at solar maximum, using the cosmic ray spectrum of \citet{kuznetsov2005models} at 1 AU (their Fig.~2, dashed line).  No shielding by a planetary magnetic field is assumed.
\item \textbf{flaring M-dwarf}: This case is based on the SCR proton spectrum measured during the September 30, 1989 solar proton event, using data measured at 1 AU by the spacecrafts GOES 6 and 7 \citep[][their Fig. 11]{smart2002review}. No shielding by a planetary magnetic field is assumed.  \label{enum:SPE}
\end{enumerate} 

For the Earth reference case, we use a solar input spectrum based on \citet{gueymard2004sun}.  The M-dwarf spectrum of AD Leo is produced from observations in the UV by the IUE satellite,  in the visible by \citet{pettersen1989spectroscopic}, and in the near-IR by \citet{leggett1996infrared}. For wavelengths longer than 2.4 $\mu$m, data from a model by \citet{hauschildt1999nextgen} was used. 
Apart from this all modelled scenarios are Earth-like as presented in Sect. \ref{sec:atm}

\begin{figure}[]
\includegraphics[width=\columnwidth]{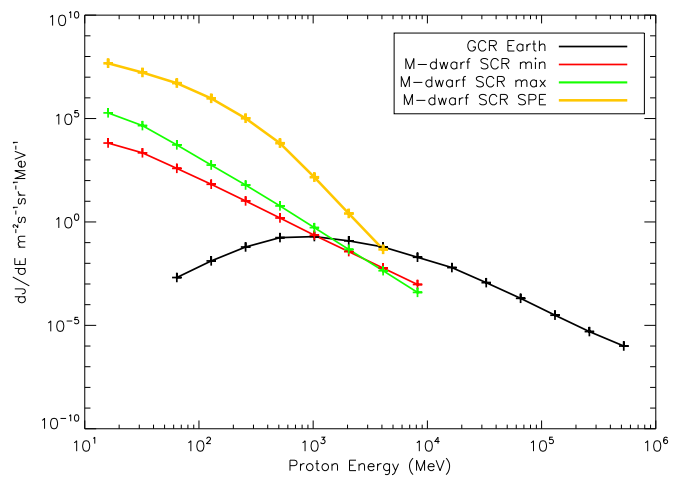} 
\caption{Cosmic ray proton flux spectra at top of atmosphere. Solar cosmic ray spectra for quiescent (case 2, red line), active (case 3, green line), and flaring (case 4, yellow line) M-dwarf scenarios (without planetary magnetic shielding) were scaled to 0.153\,AU.  The black line shows the average Galactic cosmic ray spectrum for Earth at 1 AU (case 0).}
\label{fig:scenarios}
\end{figure}

For scenarios 2, 3, and 4, we  convert the measured particle flux at 1 AU to the corresponding value the location of our test planet (0.153 AU). For this, we assume steady-state conditions, allowing us to use the scaling for the total event fluence for the instantaneous particle flux, leading to   
\begin{align}
\text{flux}(d) = \text{flux}(1 AU) * \left( \frac{d}{1 AU} \right)^{\beta}
\end{align}
We use an energy-independent value of $\beta = -2.0$ (see discussion in \citealp{grenfell2012response} for details).
The resulting cosmic ray energy spectra reaching the top-of-atmosphere of our test planet are shown in Fig.\:\ref{fig:scenarios}. 
The SCR spectra show strongly increased fluxes at lower energies and lower fluxes at higher energies compared to the GCR case for Earth. The CR flux is input into the atmosphere model to simulate different stellar activities and hence cosmic radiation environments on the modelled planet, which influence planetary climate and photochemistry.

As our model converges to steady-state, the proton fluxes for these scenarios are assumed to be constant. With this in mind, the flaring M-dwarf case described in scenario 4 requires further explanation. \citet{grenfell2012response}  assume that the frequency of flares for strongly flaring M-dwarfs is fast when compared with the chemical relaxation time scales in the middle atmosphere of Earth-like planets. That said, a constantly flaring M-dwarf shows in itself flares of highly irregular intensity. 
We take the flaring scenario as an upper limit although more observations of flaring M-dwarfs are needed.

\subsection{Atmosphere model} \label{sec:atm}

The one-dimensional, radiative-convective atmosphere model and the photochemical model used in this work were originally developed to simulate atmospheric scenarios of the early Earth \citep[see][]{kasting1984effects,kasting1985oxidant}. Both models were first coupled by \citet{segura2003ozone} to study the effect of different stellar spectra on the detectability of biosignatures on their host star. 
 We use the new version of this model, described in \citet{grenfell2012response,rauer2011potential} and references therein, to simulate an Earth-like (i.e N$_2$, O$_2$ dominated) atmosphere.  
For the Earth-like scenarios considered here we focus on responses from the troposphere up to the middle atmosphere (typically at 60-70km) since this region
is expected to be important for biosignatures such as ozone and nitrous oxide. 
The two coupled submodules are described in the following.

\subsubsection{Climate module}

The climate module calculates globally, diurnally averaged temperature, pressure and water profiles by assuming two energy transport mechanisms: radiation and convection. 
The model ranges from the surface to a pressure level of $6.6 \cdot 10^{-5}$\,bar using a variable pressure grid. 
As initial values, the bulk composition, temperature, and pressure are set to modern Earth profiles based on the U.S. Standard Atmosphere (COESA, 1976)\nocite{coesa1976us}. The surface albedo is fixed to 0.2067 and the gravitational acceleration to 9.8\,m/s$^2$. The radiative transfer scheme for the incoming stellar short-wave spectrum (ranging from 237.6\,nm to 4.545\,$\mu$m) is calculated using the $\delta$-two-stream method \citep{toon1989rapid}. The scheme incorporates visible and near-IR absorption by O$_2$, O$_3$, CH$_4$, and CO$_2$ and H$_2$O as well as Rayleigh scattering for N$_2$, O$_2$, and CO$_2$. The long-wave (3.33 – 1000 $\mu$m) transfer is handled by the Rapid Radiative Transfer Model \citep[RRTM;][]{mlawer1997radiative}, which takes continuum and gaseous opacity of H$_2$O, CO$_2$, CH$_4$, O$_3$, N$_2$O into account. Both long and short-wave schemes use the correlated-k method. In the IR, the k-coefficients are precalculated for a specific range in composition, temperature, and pressure, the validity range of the model ranges from 0.01 to 1050 mbar and within $\pm 30$\,K of the global-mean Earth temperature profile. Below the tropopause the temperature profile is determined by a wet adiabatic lapse rate, based on \citet{kasting1988runaway} and \citet{ingersoll1969runaway}, using the relative humidity profile of  \citet{manabe1967thermal}.

The climate module converges when the relative temperature difference between two adjacent iteration steps does not exceed a value of $10^{-7}$ in all levels and the total radiative flux (downwards and upwards) at top of atmosphere is less than $10^{-3}$ W/m$^2$ for a timestep larger than $10^4$ seconds. 
The module then outputs converged to steady-state solutions for the globally and diurnally-averaged temperature, pressure, and water profiles,   which are then passed on to the chemistry module.

\subsubsection{Chemistry module}

The chemistry module calculates the steady-state solution of a photochemical network of over 200 reactions of 55 species, which are important for  biomarker and greenhouse gas chemistry on a planet with an Earth-like development. The scheme thereby solves a set of non-linear continuity equations using an implicit Euler method. 
When the species' concentrations converge, loss rates (e.g. due to gas-phase in-situ sinks, condensation, deposition at the lower boundary, wet deposition, etc.) are then in balance with species' production rates (e.g. due to gas-phase in-situ sources, evaporation, emission and effusion fluxes at the model boundaries, etc.).

 We assume modern Earth values for well-mixed isoprofile species (N$_2 \approx 0.78$, O$_2 = 0.21$, Ar = 0.01 and CO$_2 = 3.55 \cdot 10^{-4}$ volume mixing ratio) and for surface biogenic fluxes of for example CH$_3$Cl, N$_2$O, and CH$_4$ \citep[see][]{grenfell2013potential}. An eddy diffusion coefficient profile \citep{massie1981stratospheric} is used to parameterise the vertical transport of species between model layers. 
The chemical scheme includes photolysis rates for the major absorbers  and more than 
200 chemical reactions including the essential chemical families Ox, HOx, NOx, and SOx, and is designed to reproduce the ozone profile on modern Earth. More details of the chemical network scheme are
provided in \citet{kasting1985oxidant},  \citet{segura2003ozone} with recent updates described in \citet{rauer2011potential}.\footnote{ Further detail on the species, absorbtion coefficients, chemical reaction network, etc. can be found in the following link http://vpl.astro.washington.edu/sci/AntiModels/models09.html for a version of the model code as described by \citet{segura2003ozone,segura2005biosignatures}.} The reaction rates are set to the recommendations of the JPL 2003 report \citep{sander2003chemical}.
Photolysis rates in the chemistry module are calculated for ten far-UV wavelength intervals from 120-175\,nm, i.e. including Ly-alpha, which is important 
to capture water photolysis in the upper atmospheric layers for example. 
The remaining 108 wavelength intervals range from  175.4 -855.0\,nm. The intensity of the photon flux is calculated with the $\delta$-two stream method \citep{toon1989rapid} for  175.4 -855.0\,nm and with Beer-Lambert's law for  120-175\,nm.

The chemistry module converges when all species concentrations in all model layers change by less than $10^{-4}$ of their relative amount between adjacent iteration steps with $dt>10^5$ seconds.
The resulting converged profiles of the radiative species are passed back to the climate module. This process iterates until the whole model converges to a steady-state solution, which is reached when the relative difference of the temperature profiles of two consecutive radiative-convective/photochemical iterations does not exceed $10^{-5}$. This is satisfied in most cases after 70-100 iterations.

\subsection{Improvements to the cosmic ray scheme} \label{sec:imp}

The original model described in \citet{grenfell2007biomarker,grenfell2012response} uses an atmospheric air-shower approach to simulate CR-induced NO production rates (which were fed into the chemistry module) from incident primary particles at TOA. The only primary particles taken into account are protons, which in turn produce only secondary electrons. 
Particles are assumed to propagate in straight lines, according to their initial incident angle. 
This approach 
is usually valid for particle energies between TeV and EeV \citep{alvarez2002hybrid}. In our model, however, we apply this approach for particle energies of MeV - TeV. Thereby we make use of extrapolated values for the Gaisser-Hillas-Function (see Sect.\:\ref{sec:gaihil}). The resulting secondary electron fluxes reproduce the measured maximum of ionizing radiation by CRs at 15$\pm$2\,km, also called the ``Pfotzer maximum" \citep{pfotzer1936dreifachkoinzidenzen}.

In the current work, we utilise an improved version of the cosmic ray scheme used by \citet{grenfell2012response}. The key updates  include 
\begin{itemize}
\item an improved CR-induced production of NOx molecules and a new CR-induced HOx production based on the ion-pair production rate (IPR) from CR particles (Sect.~\ref{sec:ipr})
\item an improved, energy-dependent impact cross-section of CR-induced secondary electrons (Sect.~\ref{sec:sigma})
\item updates to the parameters of the Gaisser-Hillas scheme used by \citet{grenfell2012response} (Sect.~\ref{sec:gaihil})
\end{itemize}

\subsubsection{NOx and HOx production via ion pair production rate} \label{sec:ipr}

The CR  scheme used in this work uses an air-shower approach based on the parametrisation by \citep{gaisser1977reliability} to derive secondary electron fluxes. In previous works \citep[e.g.][]{grenfell2007biomarker,grenfell2012response}, it was assumed that these electrons directly produce NO via
\begin{align}
\text{N}_2 + \text{e}^- &  \rightarrow 2\text{N($^4$S)} + \text{e}^- \\
\text{N($^4$S)} + \text{O}_2 &  \rightarrow \text{NO} + \text{O($^3$P)},
\end{align}
with the assumption that $k_{\text{NOx}}=1$, i.e one NO molecule would be produced on average per destroyed N$_2$ molecule. Thus \citet{grenfell2007biomarker,grenfell2012response} use an ion pair production parametrisation with a production rate of 1 NO per ion pair \citep{nicolet1975production}.

The current work improves upon this scheme by parametrically taking into account  ionic products of the CR-induced destruction of the main atmospheric constituents N$_2$ and O$_2$, adding a HOx production parametrisation, as well as incorporating branching ratios of the produced NOx and HOx species. This approach is necessary when modelling HOx production as one would otherwise have to incorporate a complex ion chemistry network involving ionic water clusters \citep[see e.g.][]{verronen2013analysis}.

We compute the IPR (ion pair production rate) by summing up the ionisation rates by electron impact for N$_2$ and O$_2$, i.e. 
\begin{align}
IPR(X) =\sum_{i=N_2,O_2} \int^{E_2}_{E_1} n_{i}(X) \cdot \sigma_{ion(i)}(E_{el}) \cdot F_{el} (X,E_{el}) dE_{el} \label{eqn:ipr}
\end{align}
where $IPR(X)$ is the ion pair production rate for the layer with mass column density $X$, $n_{\text{i}}$ is the number density of species $i$, $\sigma_{ion(i)}$ is the total ionisation cross-section of chemical species $i$ by electron impact, and $F_{el}$ the flux of secondary electrons. The integral over electron energy $E_{el}$ is confined by the energy range ($E_1$,$E_2$) of $\sigma_{ion(i)}$, i.e. $\sim$10 to 1000 eV for the current work. 
The ionisation cross-sections for N$_2$ and O$_2$ are given in Fig. \ref{fig:sigma}. 

For this approach, production coefficients $k$ are used not only for NOx production where $k_\text{NOx}=1.25$ \citep{porter1976efficiencies,jackman1980production}, but also for the production of HOx where $k_\text{HOx}=2$ \citep{solomon1981effect}.

A further improvement is that we take into account branching ratios for both NOx as well as HOx species, which are well known. 
The reaction of N$_2$ with CRs results in 45\% of ground state nitrogen   N($^4$S) and 55\% excited state nitrogen, e.g.  N($^2$D), N($^2$P) \citep{jackman2005neutral,porter1976efficiencies}. 
In accordance with \citet{jackman2005neutral}, we do not directly include N($^2$D) in our chemical scheme, but assume that the reaction
\begin{align}
\text{N($^2$D) + O$_2$ $\rightarrow$ NO + O($^3$P)}
\end{align}
proceeds instantly 
 while ground state N can either produce or destroy NO via 
\begin{align}
\text{N($^4$S) + O$_2$} &  \rightarrow \text{NO + O($^3$P)} \label{eqn:NO1}\\ 
\text{N($^4$S) + NO}      &  \rightarrow \text{N$_2$ + O($^3$P)} \label{eqn:NO2},
\end{align}
depending on the pressure and temperature of the surrounding medium. 

The parameter $k_\text{HOx}$ strongly depends on the amount of available atmospheric H$_2$O. For the purposes of the current work, we assume the maximum value $k_\text{HOx}=2$ as we are interested in altitude regions below 60 km (compare with Fig. 2  in \citet{solomon1981effect}) and our model atmospheres provide sufficient H$_2$O concentrations in that region.
For HOx a production coefficient $k_\text{HOx}$ of 2 HOx molecules per ion pair produced is equivalent to the destruction of a water molecule (with $k_\text{H}=k_\text{OH}=1$):
\begin{align}
\text{H$_2$O} + \text{Ion}^+ \rightarrow \text{H} + \text{OH} + \text{Ion}^+
\end{align}
Recent studies by \citet{sinnhuber2012energetic} have found that $k_\text{OH}=0.9$ in the mesosphere, yet testing this change (not shown) had little to no effect on our results.

\subsubsection{Energy dependent electron-impact cross-sections} \label{sec:sigma}

The above mentioned additions to our rather simple air-shower scheme provide a motivation to improve our model input parameters in the CR scheme. The electron impact cross-section $\sigma$ is one such parameter.
Our previous studies \citep{grenfell2007biomarker,grenfell2012response} have used a constant value for the electron impact cross-section for N$_2$ of $\sigma_{N_2} = 1.75 \cdot 10^{-16}$\,cm$^2$ \citep{nicolet1975production} for electron energies between 30eV and 300eV (and $\sigma_{O_2} = 0$). The current work uses energy dependent total ionisation cross-sections $\sigma_{ion(N_2)}$, $\sigma_{ion(O_2)}$,  which are needed to calculate the ion pair production rate via Eq.~\ref{eqn:ipr}.

\begin{figure}[h]
\includegraphics[width=\columnwidth]{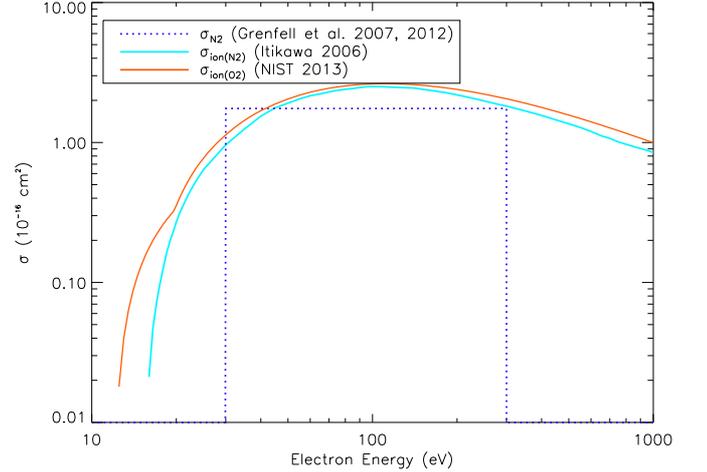} 
\caption{Comparison of previously used cross-section $\sigma_{N_2}= 1.75 \cdot 10^{-16}$\,cm$^2$ and new total ionisation cross-sections $\sigma_{ion(N_2)}$ \citep{itikawa2006cross} and $\sigma_{ion(O_2)}$ \citep{NIST2013}.}
\label{fig:sigma}
\end{figure}

Figure \ref{fig:sigma} shows both the old, constant value of $\sigma_{N_2}$ as well as our updated N$_2$ and O$_2$ ionisation cross-sections.  Using the new values  better reflects the contribution of the involved particles according to their energies.   The overall impact of these and all further changes is shown in Sect.\:\ref{sec:val}. 

\subsubsection{Expanding energy range and improving energy dependence of the input parameters of the Gaisser-Hillas scheme} \label{sec:gaihil}

 Since \citet{griessmeier2014} work with primary protons with an extended energy range of 16\,MeV to 0.5\,TeV (compared to the previously-used range of 64\,MeV to 8\,GeV), the parameters of the Gaisser-Hillas formula need to be extended as well. The scheme is detailed by \citet{grenfell2007biomarker}, but we nevertheless give a short review here.

We wish to calculate the secondary electron flux $F_{el}$ from the top-of-atmosphere proton flux, so that $F_{el}$ can then be input into Eq.\:(\ref{eqn:ipr}). 
At first 
$F_{el}$ is separated into
\begin{align}
F_{el}(X,E_e) = R_{el}(X) \cdot S_{el}(E_e) \label{eqn:Fel},
\end{align}
the total electron flux $R_{el}$ and the spectral distribution of electrons $S_{el}(E_e)$, which is assumed to follow a power law \citep{grenfell2007biomarker} 
\begin{align}
S_{el}(E_e) = a \cdot E_e^{-3} \label{eqn:Sa}
\end{align}
with the criterion that $S_{el}(E_e)$ is normalised over the whole energy range
\begin{align}
\int S_{el}(E_e) d E_e = 1.
\end{align}
 As we see during the validation in Sect.\:\ref{sec:val}, measurements of the resulting NO production rate are best reproduced when the parameter $a$ in Eq.\:(\ref{eqn:Sa})  is set to a value of 18 eV$^2$.

The total electron flux $R_{el}$ from equation (\ref{eqn:Fel}) is obtained by
\begin{align}
R_{el}(X) =  \int_\Omega d\Omega \int_{E_p}  N_{el,p}(X,\theta,\phi,E_p) \cdot F_p(X=0,E_p) dE_p,
\end{align}
where $\Omega$ is the solid angle and $N_{el,p}$ is the number of electrons produced by one primary proton with  energy $E_p$ and an incident direction ($\theta$,$\phi$) \citep{grenfell2007biomarker}. 

We obtain $N_{el,p}$ using the Gaisser-Hillas formula
\begin{align}
 N_{el,p}(X',E_p) & =   N_{max}(E_p) \cdot e^{\frac{X_{max}(E_p)-X' }{\lambda(E_p)} } \nonumber \\
&\;\;\;\;  \cdot \left(\frac{X' -X_0(E_p)}{X_{max}(E_p) - X_0(E_p)}\right)^{\frac{X_{max}(E_p)-X_0(E_p) }{\lambda(E_p)} }  \label{eqn:gaihil},
\end{align}
 which was as first introduced as a fit function for high-energy cosmic ray measurements \citep{gaisser1977reliability}.  
The equation yields the number of secondary electrons produced  $N_{el,p}$, dependent on the mass column density $X'$ in the trajectory of the primary particle and its energy $E_p$. The column density  along a trajectory $X'$ at a given height $z$ depends on the incident angle $\theta$ of the proton and the mass column density at $z$, 
\begin{align}
X'(z) = \frac{X(z)} {\cos(\theta)}. \label{eqn:costheta}
\end{align}
The free parameters of Eq.~(\ref{eqn:gaihil}) are the position $X_{max}$ of the electron maximum, the number $N_{max}$ of secondary electrons at $X_{max}$, the proton attenuation length $X_0$, and the electron attenuation length $\lambda = 40$, where $X_{max}$, $X_0$, and $\lambda$ are each in units of mass column density, i.e.~g/cm$^2$. Table \ref{tab:gaihilpara} lists the parameters used by \citet{grenfell2012response}. We now address each of these parameters in turn.

\setlength{\tabcolsep}{2.2pt}
\begin{table}[]
\centering
\begin{tabular}{rcccccccc}
$E_p$ [MeV]  & 64  & 128  & 256 & 512 & 1024 & 2048 & 4096 & 8192 \\
\hline
\hline
$X_0$ [g/cm$^2$]                    & 0.001& 1   & 2     &   80 &   80   &   80   &     80 &     80  \\
$X_{max}$ [g/cm$^2$]            & 0.01& 5     & 25   & 150 & 150   & 150   &   150 &   150 \\
$N_{max}$           & 3 & 3 & 3 & 3 & 3 & 3 & 3 & 3 \\
\end{tabular}
\caption{Parameters of the Gaisser-Hillas function (Eq.~\ref{eqn:gaihil}) following \citet{grenfell2012response}}
\label{tab:gaihilpara}
\end{table}
\setlength{\tabcolsep}{3pt}

\paragraph{Proton attenuation length $X_0$:}

For this parameter we adopt the air-shower approach of  \citet{alvarez2002hybrid}, who also use the Gaisser-Hillas function, but assume $X_0=5$ g/cm$^2$ as the depth of the first interaction. We therefore set  $X_0$ to 5\,g/cm$^2$ for proton energies equal to 512\,MeV and above. This has the effect that the peak of our resulting secondary electron flux, previously situated rather sharply at an altitude of about 15\,km for Earth, spreads over a wider region, thereby corresponding better to Earth measurements (see Sect.\:\ref{sec:val} and Fig.~\ref{fig:NOcomp}).

For energies below 512\,MeV, the air-shower scheme applied here  loses accuracy as it is better applied to high-energy primary particles. This outcome is most easily seen, when $X_0$ > $X_{max}$, as this would produce negative fluxes in Eq.~(\ref{eqn:gaihil}). Furthermore, primary protons of such low energy are normally not involved in particle cascades or air-showers, but rather lose energy directly to the atmosphere via ionisation \citep{usoskin2009ionization}. To address this problem, we lower the values of $X_0$ for primary protons with energies below 512\,MeV to the values shown in Table \ref{tab:lowen} and \ref{tab:newgaihil}. This means these protons interact with the atmosphere at higher altitudes in our model, which acts as an approximation of their energy loss via ionisation of the upper atmosphere.

\paragraph{Position of the electron maximum $X_{max}$:}

This parameter is strongly energy dependent, so it needs to be addressed accordingly  when expanding the energy range. In their Fig.~1, \citet{swordy2000elemental} give theoretical values of $X_{max}$ for incident protons from 100\,TeV to 10\,PeV.

\begin{figure}[]
\includegraphics[width=\columnwidth]{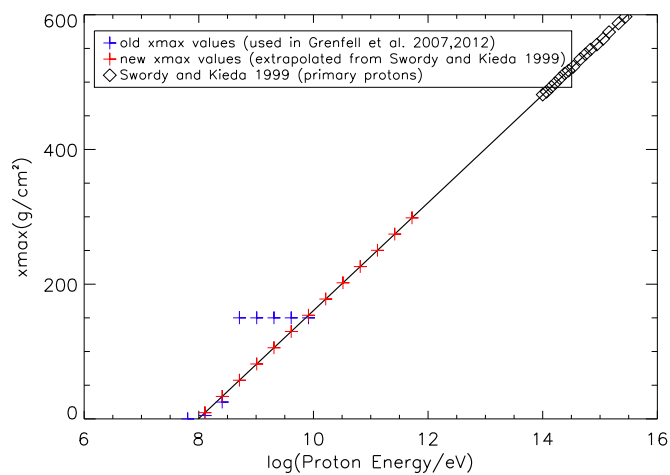} 
\caption{Values for $X_{max}$ from \citet{swordy2000elemental} (black diamonds). We extrapolated the values to 64\,MeV - 0.5\,TeV (red crosses) for the current study. Also shown are the values for $X_{max}$ as used by \citet{grenfell2012response} (blue crosses).} 
\label{fig:swordy}
\end{figure}

Figure \ref{fig:swordy} shows an extrapolation of the values obtained from \citet{swordy2000elemental} to the energy range relevant to this work. The extrapolated values take  the expected energy dependence of $X_{max}$ into account and show reasonable agreement with the previously used values at 100\,MeV and 10\,GeV.

In the previous CR scheme, the effect of CR protons with energies lower   
 than 64\,MeV were not included.  
These may, however, play an important role for the simulation of the effect of SCRs upon the atmosphere, as SCR proton spectra feature many primary protons with energies lower than 64\,MeV \citep{wissing2009atmospheric} (see also Fig.~\ref{fig:scenarios}). 
As suggested above, the range of validity of our model reaches its limit at low energies. Nevertheless, a rough assumption is made here for energies between 16 - 64\,MeV. Monte-Carlo simulations of downwards proton transfer in the atmosphere performed by \citet{wissing2009atmospheric}  (their Fig.~2) suggest  values for $X_{max}$ in the stated energy range, which are given in Table \ref{tab:lowen}. We assume the proton attenuation length $X_0$ to scale accordingly. 

\begin{table}[h]
\centering
\begin{tabular}{r c c c}
$E$ [MeV]    & 16 & 32 & 64  \\
\hline
\hline
$X_{max}$  & 0.01 & 0.65 & 1.7\\
$X_0$          & 0.001& 0.01   & 0.1 \\
\end{tabular}
\caption{Parameters for lower energy ranges, adapted from \citet{wissing2009atmospheric}.}
\label{tab:lowen}
\end{table}

\paragraph{Maximum number of electrons $N_{max}$:}

The maximum number of electrons produced per proton (at $X_{max}$) is also highly energy dependent, but was used as a freely-adjustable factor in previous works. \citet{alvarez2002hybrid} provide an analytical relation, with a validity range of $10^{14} - 10^{19}$ eV 
\begin{align}
N_{max} = 0.75 \cdot \frac{E_p}{\text{GeV}}\label{eqn:nmax}
\end{align}
As with the previous parameters, we also extrapolate this relation to the energy range relevant for this work (see Table \ref{tab:newgaihil}).

\begin{table*}
\centering
\begin{tabular}{r  c c c c c c c c}
$E$ [MeV]    &16      &32     & 64  & 128  & 256        & 512 & 1024 & 2048 \\
\hline
\hline
$X_{0}$       & 0.001 & 0.01 & 0.1&   1     & 2        &   5    &   5      &     5          \\
$X_{max}$   & 0.01 & 0.65 & 1.7 & 9.2     & 33.3   & 57.4 & 81.5   & 105.6   \\
$N_{max}$  & 0.125 & 0.25 & 0.05 &0.10 & 0.19 & 0.38 & 0.77   &1.54    \\\\
$E$ [MeV]& 4096  & 8192 & 16384 & 32768 & 65536 & 131072 & 262144 & 524288 \\
\hline
\hline
$X_{0}$     &     5      & 5        & 5         & 5        &        5 &     5     &  5       &   5      \\
$X_{max}$ &   129.7  & 153.8 & 178.0 & 202.0 & 226.2 & 250.3 &274.4 & 298.5 \\
$N_{max}$  & 3.07    & 6.14  & 12.29& 24.58 & 49.15  & 98.30 & 196.61 & 393.22\\

\end{tabular}
\caption{Updated set of Gaisser-Hillas parameters with improved energy dependence and expanded energy range.}
\label{tab:newgaihil}
\end{table*}

\subsubsection{Model validation} \label{sec:val}

\begin{figure}[]
\includegraphics[width=\columnwidth]{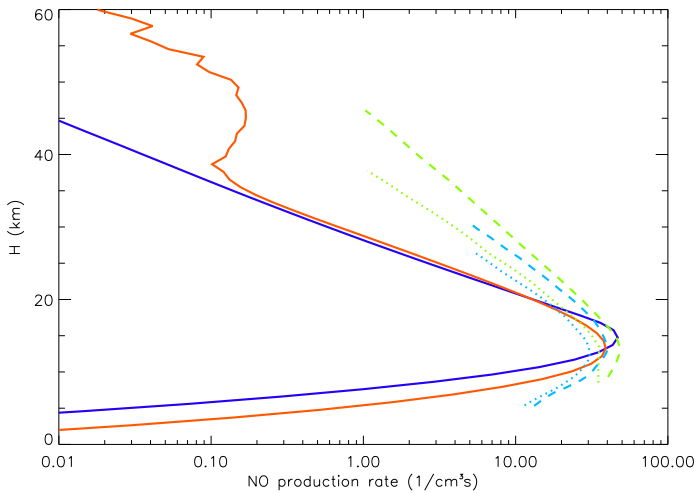} 
\caption{Comparison of GCR-induced NO production rates, showing model output for the original scheme \citep[][ dark blue solid curve]{grenfell2012response} and the new scheme (red solid curve) as well as Earth measurements from \citet{nicolet1975production} for solar maximum (blue dotted curve) and solar minimum (blue dashed curve) and from \citet{jackman1980production} for solar maximum (green dotted curve) and solar minimum (green dashed curve).}
\label{fig:NOcomp}
\end{figure}

The different model components have been validated as follows:

\noindent\underline{GAS-PHASE CHEMISTRY} - \citet{grenfell2013potential} showed that the model (without CRs) calculates a  (global, annual mean) 
ozone column of 305 Dobson Units is in agreement with modern Earth observations. That work also showed that the Chapman mechanism  producing ozone dominates in the model stratosphere, peaking 
near the observed ozone maximum, and that the so-called smog mechanism dominates in the troposphere. These findings are in-line with current  understanding of ozone photochemistry 
on the modern Earth.  That work also presented column values for other (in addition to ozone) biosignature candidates, related source  gases as well as surface UV 
which suggested that the model reproduces the modern Earth reasonably well.  \citet{grenfell2011sensitivity} found reasonable  agreement between 
the surface biomass fluxes required in the chemistry module to reproduce the modern Earth atmospheric composition and the observed  values for these fluxes. 

\noindent\underline{RADIATIVE TRANSFER} -  \citet{rauer2011potential} showed temperature profiles for the modern Earth which  suggested that the model reproduces observations of the modern 
Earth reasonably well. 

\noindent\underline{THIS WORK,  COSMIC RAY SCHEME}  -  SCRs are difficult to validate with Earth observation using our stationary (i.e. without time dependence) model because they are a time-dependent phenomenon 
so we here focus only on GCRs. There are three points to consider:

Firstly, the effects of secondary electrons produced from air-shower schemes are commonly validated by comparing the computed ion pair production rates (IPR)
with those observed.   Our model-calculated  IPR (which is the NO production of Fig.~\ref{fig:NOcomp} rate divided by $k_\text{NOx}=1.25$, see Sect.~\ref{sec:ipr}) falls within the range calculated by \citet{velinov2008improved} between solar minimum and maximum in the middle stratosphere i.e. 3 to 30 ion pairs per cm$^{-3}$ s$^{-1}$, depending on solar cycle phase and altitude.

Secondly, the high-energy secondary electrons break up the strongly-bound N$_2$ molecule, and the resulting nitrogen atoms quickly react with oxygen-containing species
which leads overall to an enhanced production rate of NO (see Fig.~\ref{fig:NOcomp}). 
We compare our modelled production rate of nitrous oxide (NO) with data derived from measurements by \citet{nicolet1975production} and \citet{jackman1980production}. 
Using the improvements described in Sect.\:\ref{sec:gaihil}, we calculate the NO production rate as shown in 
Fig.~\ref{fig:NOcomp}. 
The blue solid curve used the original scheme, while the solid red curve used the improvements presented in sections  \ref{sec:ipr} -- \ref{sec:gaihil}. The remaining curves show data obtained from measurements by \citet{nicolet1975production} and \citet{jackman1980production}, respectively, for minimum and maximum solar activity at 60$^\circ$ geomagnetic latitude.
We see an improvement when comparing the NO production rates of both the current and previous work. 
Overall the new scheme produces a broader peak below 20\,km, which is shifted down in altitude and whose maximum is slightly reduced, thereby producing better agreement with the measured rates. This effect is especially evident at altitudes below 13 km, which are due to the incorporation of higher energy incident protons. 
These higher energy protons affect lower altitudes due to increasing $X_{max}$ values with proton energy (see Sect.~\ref{sec:gaihil}). Additionally,  high-energy protons have a greater effect on the atmosphere due to higher values of $N_{max}$ for higher energies. The peak NO production for the new scheme is broadened because of the change in $X_0$. The parameter $a$ (see Eq.~\ref{eqn:Sa}) was set to 18\,eV$^2$ to move the peak value into the range of measured NO production rates. Above 35 km the effects of the new IPR scheme (see Sect.\:\ref{sec:ipr}) are particularly evident. In this region the branching between NO-producing and NO-destroying effects of ground-state nitrogen (see Eqs.\:\ref{eqn:NO1} + \ref{eqn:NO2}) favours NO production. The curve is slightly jagged because of the non-linear dependence of Eqs.\:(\ref{eqn:NO1}) and (\ref{eqn:NO2}) upon 
 pressure, temperature and concentration. 
In the altitude region between 20 and 35 km the model results underestimate the observations. While adding a background SCR spectrum to the incident protons could mitigate this discrepancy to some extent, the most probable cause is that we need to better address the gradual loss of proton energy via ionisation of the atmosphere. 
This issue is beyond the scope of the current work. 
 Nevertheless, we are  able to reproduce the effects of the Pfotzer maximum using both old and new schemes, and both modelled curves show good agreement to the data obtained from measurements in this region. However, the new scheme works with ionic secondary particles so that NOx and HOx production parametrizations can be calculated via the IPR approach. In addition the energy ranges for primary particles were extended from 64\,MeV - 8192\,MeV to 16\,MeV - 524288\,MeV and our new scheme uses updated input parameters that are energy dependent, which allows for primary particles to produce a different number of secondaries depending on their initial energy.  

Thirdly, the model calculates a concentration of NO based on its sources and sinks (including the GCR source).
The modern Earth scenario (including GCRs) produces a concentration of nitrogen monoxide (NO) of 12\,ppbv at 40\,km 
which is consistent with modern Earth observations \citep[see e.g.][]{sen1998measurements}. However, this species varies both diurnally and over latitude so a 
clear consensus of its global average (i.e. as calculated in our model) is not available.

While it would be desirable to validate directly with OH production, OH is a difficult species to observe especially in the stratosphere. OH is a highly reactive
trace species and its stratospheric mean value is not well known. Faced with the lacking observations, some recent 3D model studies from the Earth science community have
been recently applied to estimate the effect of GCRs on HOx.  For example, \citet{calisto2011influence} reported HOx decreases due to GCRs of 1-3\% in the stratosphere. Our model calculates a column response of 3\% decrease of HOx (H + OH + HO$_2$) due to CRs. While these values are broadly comparable they cannot be directly compared to the 3D model study of \citet{calisto2011influence}  as that study does not provide global mean values. In a similar fashion the response of O$_3$ concentrations can be mentioned. Our model calculates an integrated column density loss of O$_3$ of less than 1\% due to GCRs on Earth, while \citet{calisto2011influence} find that GCR cause  up to 3\% loss of O$_3$ in the polar stratosphere and upper troposphere as well as an increase in O$_3$ of up to 3\% in the tropics and south polar troposphere.

\subsection{Line-by-line spectral model} \label{sec:spec}

We use a radiative transfer line-by-line tool called SQuIRRL (Schwarzschild Quadrature InfraRed Radiation Line-by-line) \citep{schreier2001a,schreier2003mirart}. This tool uses a radiative transfer method with line-by-line absorption from the HITRAN 2004 database \citep{rothman2005hitran}  to compute planetary spectra of H$_2$O, CO$_2$, O$_3$, CH$_4$, N$_2$O, CO, SO$_2$, NO$_2$, NO, HCl, HNO$_3$, H$_2$O$_2$, HO$_2$, CH$_3$Cl, OH, ClO, and ClONO$_2$. The converged density profiles of the above mentioned species, as well as pressure and temperature from the atmospheric column model are input into SQuIRRL.

\section{Results}

\subsection{Atmospheric profiles} \label{sec:ap}

\subsubsection{Ion pair production rates} \label{sec:ip}

\begin{figure}[]
\includegraphics[width=\columnwidth]{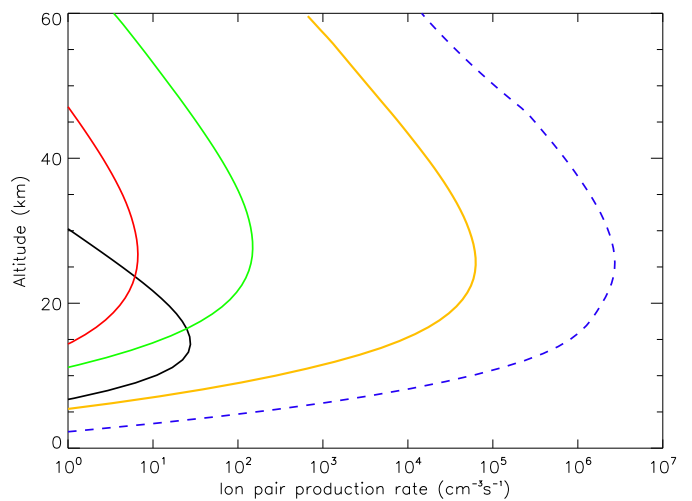} 
\caption{Ion pair production rates as output by the CR air-shower scheme. Shown are: scenario 0 (GCR Earth, black curve), scenario 2 (quiescent M-dwarf, red curve), scenario 3 (active M-dwarf, green curve), scenario 4 (flaring M-dwarf, yellow curve), as well as the flaring case from \citet{grenfell2012response} (dashed blue curve, scenario 6).}
\label{fig:iprcomp}
\end{figure}

Figure \ref{fig:iprcomp} shows the model output ion pair production rates (see Eq.~\ref{eqn:ipr}) for the incident proton spectra shown in Fig.~\ref{fig:scenarios}. Solid lines were calculated using the new scheme, while the dashed line originates from the scheme as used by  
\citet[][]{grenfell2012response} in their flaring case.   %here
The modelled IPR of GCRs (black line) and SCRs (coloured lines) in Fig.~\ref{fig:iprcomp} peak at decidedly different altitudes, with GCRs peaking at around 14\,km and SCRs at 25 to 30\,km altitude. While the IPR is relatively small for the quiescent M-dwarf case (red line), the IPR intensity increases significantly with activity of the host star (green and yellow lines). 

Noteworthy is the change in IPR intensity when comparing the flaring M-dwarf scenario of this work with that of \citet{grenfell2012response}. The new scheme results in an IPR at its maximum that is roughly 40 times smaller than the IPR produced by \citet{grenfell2012response}. 
This significant change in the effect of CRs is a direct result of our updated input parameters (see Sect.\:\ref{sec:gaihil}). In particular, the expansion of the energy range of the Earth GCR validation run with energies $>8.2$\,GeV, together with $N_{max}(E)$ \textasciitilde E, significantly increases the contribution of CR particles with $E>8.2$\,GeV to the validation spectrum. The SCR fluxes of these high energies are, however, negligibly small when compared to the maximal SCR flux. Hence the overall SCR-induced IPR profiles decrease because the influence of $E>8.2$\,GeV GCR protons is increased.
While  the updated scheme provides a more physically valid parametrisation of the effect of cosmic ray ionisation (especially regarding the energy dependence of the parameters), we nevertheless wish to compare the chemical effects of our improved scheme with those of the original. For this purpose we include two new scenarios in our analysis as shown below:

\begin{enumerate} 
\setcounter{enumi}{4}
\item \textbf{flaring(x40) M-dwarf}: same as scenario \ref{enum:SPE}, but with 40 times enhanced IPR.
\item \textbf{flaring M-dwarf \citep{grenfell2012response}}: same as scenario \ref{enum:SPE}, but with the original version of the CR scheme as described by  \citet{grenfell2012response}.
\end{enumerate}

\subsubsection{Chemistry profiles} \label{sec:chemm}

\begin{figure}[]
%\centering
\includegraphics[width=\columnwidth]{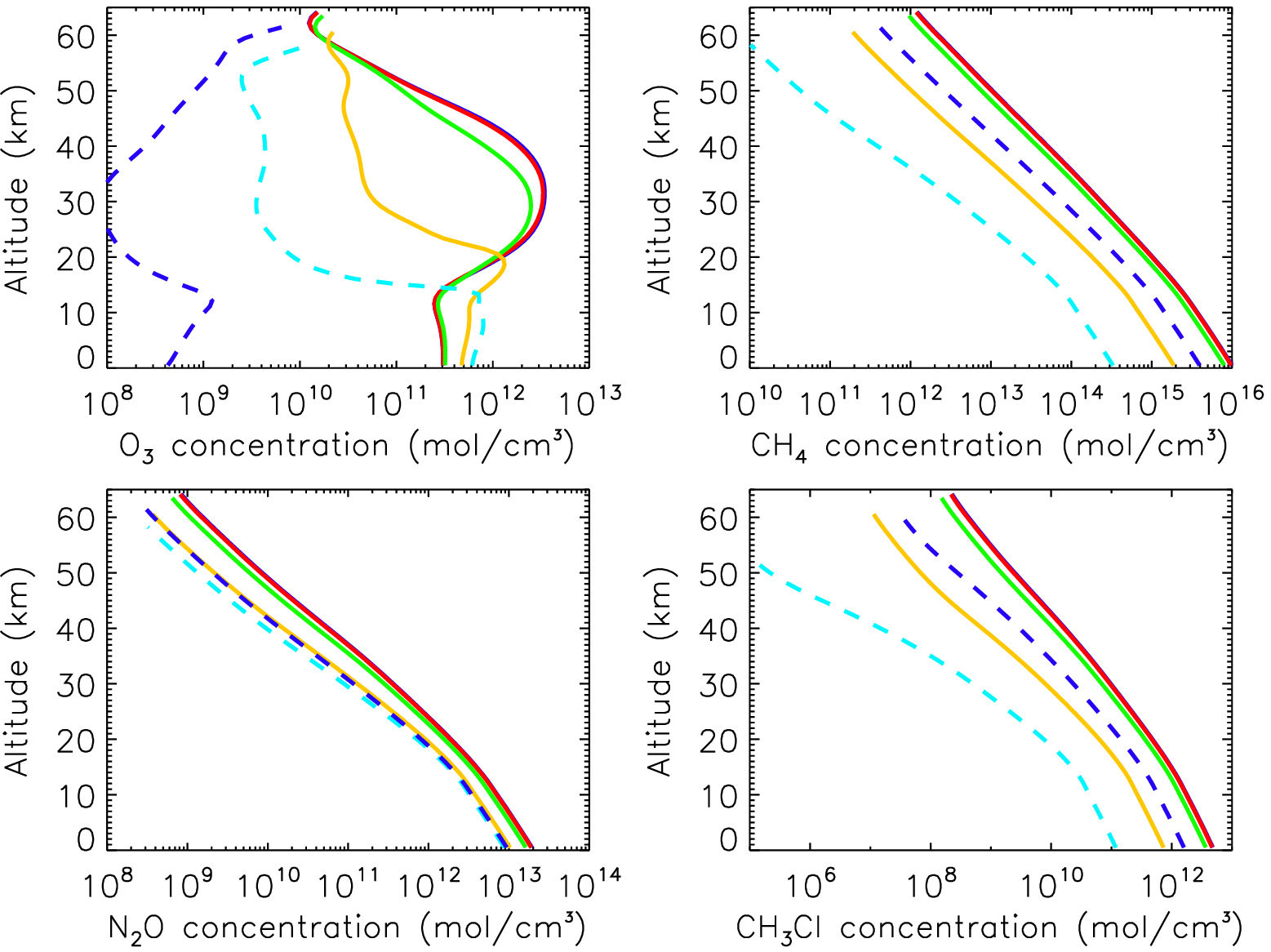} 
\includegraphics[width=\columnwidth]{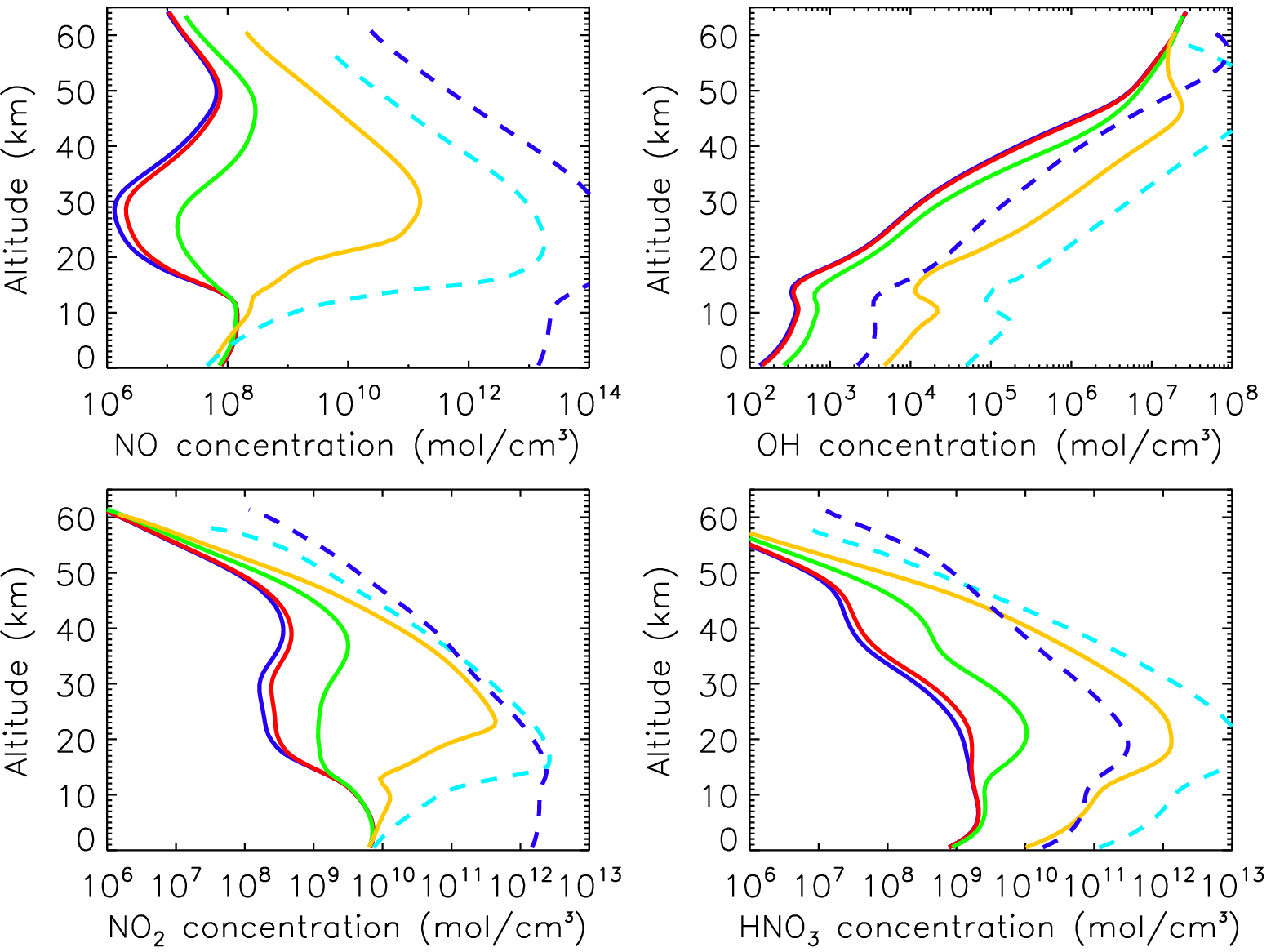} 
\caption{Chemical concentration profiles for biomarker species O$_3$, CH$_4$, N$_2$O, and CH$_3$Cl and for related species NO, OH, NO$_2$, and HNO$_3$ in molecules per cm$^3$. Shown here are scenarios 1 through 5: M-dwarf w/o CR (solid dark blue line), quiescent M-dwarf (red line), active M-dwarf (green line), flaring M-dwarf (yellow line), flaring(x40) M-dwarf (light blue dashed line). In addition we show the result of \citet{grenfell2012response}'s flaring case (dark blue dashed line) for comparison.  }
\label{fig:chem}
\end{figure}

Figure \ref{fig:chem} shows chemical profiles for the biomarker species O$_3$, CH$_4$, N$_2$O, and CH$_3$Cl as well as the related species NO, OH, NO$_2$ and HNO$_3$ for scenarios 1 through 6. The M-dwarf w/o CR (blue line) is very similar to the AD Leo case of \citet{rauer2011potential}, as we use the same version of the atmospheric model. Increasing the CR intensity has noticable effects on the modelled atmospheres. While only small changes in the NO, NO$_2$ and HNO$_3$ concentrations are observed for the quiescent M-dwarf, the active and flaring scenarios (scenarios 3 and 4) show significant variations. The active case shows a column-integrated decrease in O$_3$ by 30\% and CH$_4$ by 15\% as well as a 300\% increase in HNO$_3$ compared to scenario 1 (w/o CR). The flaring scenario (4) shows a column-integrated O$_3$ decrease by over 70\% and  CH$_4$ by 80\% and a significant increase of HNO$_3$ by a factor of 315.
A detailed list of column-integrated concentrations of key species can be found in Table \ref{tab:chem}.
In general CRs lead to the photochemical destruction of O$_3$, CH$_4$, N$_2$O, and CH$_3$Cl. 
 Because of the 
large amount of CR-induced NOx and HOx, these species are increasingly stored as HNO$_3$ with increasing CR intensity.

When comparing scenarios 5 and 6, the effects of the newly introduced CR-induced NOx and HOx production is clearly evident (as both scenarios calculate a similar IPR). Scenario 6 
calculates a reduction in O$_3$ by 99.99\%
because of the excessive amount of NO produced which destroys O$_3$ catalytically in the middle atmosphere. In scenario 5 this effect is mitigated by  the additionally produced OH, which converts NOx into HNO$_3$, mainly via 
\begin{align}
\text{NO$_2$ + OH + M } & \rightarrow \text{ HNO$_3$ + M}  .
\end{align}
Without CRs (case 1) the modelling of Earth-like planets orbiting in the HZ of M-dwarfs results in up to 300 times as much CH$_4$ compared to the reference case \citep[see also][]{segura2005biosignatures,rauer2011potential}. This  so-called massive CH$_4$ greenhouse effect can strongly perturb climate and habitability. However, when adding CRs including the updated HOx-production scheme (cases 2,3,4,5), this CH$_4$ is more efficiently destroyed, as the produced OH acts as a strong CH$_4$ sink via
\begin{align}
\text{CH$_4$ + OH } & \rightarrow \text{ CH$_3$ + H$_2$O} \,, \label{eqn:CH4OH}
\end{align}
so that for the flaring M-dwarf (case 4) only 50 times as much CH$_4$ is available compared to the Earth reference case.
A similar mechanism is responsible for the large decrease in CH$_3$Cl. N$_2$O concentrations, on the other hand, are not greatly influenced by the changes to the scheme. 
Overall, this shows that the CR-induced HOx chemistry and the nitrogen radical production introduced to this work has an important  effect and may not be neglected in scenarios with strong SCR flux.
\begin{table*}[]
\centering
\begin{tabular}{rlrrrr}
\# & Scenario  & O$_3$ [DU]  & CH$_4$ [DU]  & N$_2$O [DU] & CH$_3$Cl [DU]\\ 
\hline
\hline
0 & Earth reference &  306 & 1.2$\cdot 10^3$  & 233 & 0.35 \\
1 & M-dwarf w/o CR & 254 & 3.3$\cdot 10^5$  & 630 &160  \\
2 & quiescent M-dwarf &243 & 3.2$\cdot 10^5$  & 620 & 150 \\
3 & active M-dwarf & 175 & 2.7$\cdot 10^5$  & 530 & 120 \\
4 & flaring M-dwarf &71   &  6.5$\cdot 10^4$ & 350  & 24 \\
5 &flaring(x40) M-dwarf & 40   &  1.1$\cdot 10^4$ & 300 &  4\\
6 & flaring M-dwarf \citep{grenfell2012response} &0.2  &  1.3$\cdot 10^5$ & 310 & 53\\\\
\end{tabular}

\begin{tabular}{rlrrrr}
\# &Scenario  &  NO [DU] & OH [DU] & NO$_2$ [DU] & HNO$_3$ [DU] \\
\hline
\hline
0 & Earth reference&  1.4 $\cdot 10^{-1}$  &  2.1$\cdot 10^{-3}$ &  0.2 & 0.6\\
1 & M-dwarf w/o CR&  9.8$\cdot 10^{-3}$ & 8.4$\cdot 10^{-4}$ & 0.3 & 0.2  \\
2 & quiescent M-dwarf& 1.0$\cdot 10^{-2}$ &  8.4$\cdot 10^{-4}$ & 0.3 & 0.2  \\
3 & active M-dwarf& 2.3$\cdot 10^{-2}$ &  8.9$\cdot 10^{-4}$ & 0.5  & 0.6 \\
4 & flaring M-dwarf& 7.3$\cdot 10^0$ &      1.5$\cdot 10^{-3}$ &  16  &  63\\
5 &flaring(x40) M-dwarf& 8.7$\cdot 10^2$ &       1.0$\cdot 10^{-2}$ & 100 &  600\\
6 & flaring M-dwarf \citep{grenfell2012response}& 1.6$\cdot 10^4$ &       3.1$\cdot 10^{-3}$ &  170&  16\\
\end{tabular}
\caption{Column densities of O$_3$, CH$_4$, N$_2$O, CH$_3$Cl, NO, OH, NO$_2$, and HNO$_3$ in Dobson units ($1 \text{DU} = 2.69\cdot10^{20}$\,molecules/m$^2$) for scenarios 0 through 6.}
\label{tab:chem}
\end{table*}

\subsubsection{Temperature and water profiles}

\begin{figure}[]
\includegraphics[width=\columnwidth]{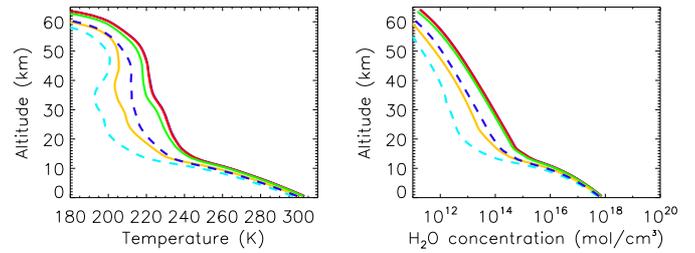} 
\caption{Temperature and H$_2$O concentration profiles for scenarios 1-6. Colouring scheme same as above.} 
\label{fig:temp}
\end{figure}

The planetary temperature profile for the scenarios with the M-dwarf spectrum we used usually shows no stratospheric temperature inversion, since the stellar UV flux is too weak to warm the stratosphere. Figure \ref{fig:temp} shows that with increasing cosmic radiation the temperature in this region generally cools by up to 30\,K owing to the loss of key radiative gases (e.g. CH$_4$, O$_3$). Since CH$_4$ abundances are
significantly higher than on Earth (by factors between 10 for scenario 5 and 300 for scenario 1; see Table \ref{tab:chem}) there is a potentially significant methane greenhouse as pointed out in previous works \citep[see e.g.][]{segura2005biosignatures,grenfell2007biomarker}.

The effect  on temperature of using the new NOx and HOx production scheme is evident from comparing the flaring scenarios 4 (yellow curve) and 6 (dark blue dashed curve): between 10 and 50\,km altitude temperatures curves of scenarios using the new scheme are markedly lower. This decrease in temperature follows from decreased CH$_4$ concentrations (see Sect.\:\ref{sec:chemm} and Fig.~\ref{fig:chem}).  H$_2$O concentrations relate well with CH$_4$, which is expected as they are highly interdependent via Eq.~(\ref{eqn:CH4OH}).

\subsection{Identifying significant chemical mechanisms} \label{sec:SPE2}

\begin{figure}[]
\includegraphics[width=\columnwidth]{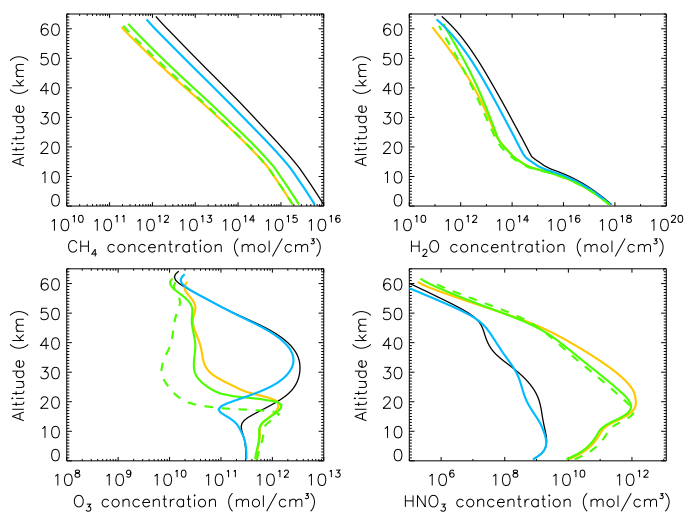}
\caption{Chemical concentration profiles for species CH$_4$, H$_2$O, O$_3$ and HNO$_3$ in molecules per cm$^3$ for scenarios 4.0 - 4.3 (see Sect.\:\ref{sec:SPE2}) with the following CR-induced production mechanisms switched on: NOx+HOx (scenario 4.0, solid yellow curve), NOx (scenario 4.1, solid green curve), HOx (scenario 4.2, solid blue curve), NO (scenario 4.3, dashed green curve). Scenario 1 (M-dwarf w/o CR) is shown for comparison (solid black curve). }
\label{fig:chemspe}
\end{figure}

To further identify important chemical mechanisms caused by the update of our CR scheme we have repeated runs of the flaring M-dwarf scenario (scenario 4), but this time with different CR-induced production mechanisms turned on. We investigate the following cases:
\renewcommand{\labelenumi}{4.\arabic{enumi}. }
\begin{enumerate} 
\setcounter{enumi}{-1}
\item \textbf{flaring M-dwarf (NOx+HOx)}: same as \ref{enum:SPE}, hence producing CR-induced NO, N, OH, and H.  \label{enum:SPE2}
\item  \textbf{flaring M-dwarf (NOx)}: same as \ref{enum:SPE}, but only produces NO and N due to CR.
\item  \textbf{flaring M-dwarf (HOx)}: same as \ref{enum:SPE}, but only produces H and OH due to CR.
\item  \textbf{flaring M-dwarf (NO)}: same as \ref{enum:SPE}, but only produces NO  due to CR, i.e. as in \citet{grenfell2012response} and otherwise using the updated Gaisser-Hillas scheme (see Eq.~\ref{eqn:gaihil}).
\end{enumerate} 
Figure \ref{fig:chemspe} shows the effects of different CR-induced production mechanisms. When comparing scenarios 4.3 and 4.1, one can infer that the O$_3$ destroying species NO is hampered in its catalytic destructivity in scenario 4.1 by CR-produced atomic N radicals, as significantly more O$_3$ is observable here. This  is caused by a self-regulating mechanism imposed by the CR-induced production of nitrogen radicals (N($^4$S)), as these can easily reconfigure into N$_2$, via e.g.
\begin{align}
\text{NO +  N($^4$S) } & \rightarrow  \text{N$_2$ + O}\,.
\end{align}
Another interesting mechanism is evident when comparing the O$_3$ concentrations for scenarios 4.0, 4.1, and 4.2. 
Both the HOx (scenario 4.2) as well as the NOx (scenario 4.1) producing schemes, lead to a decrease in O$_3$, in each case via catalytic destruction with HOx and NOx species, respectively. The NOx scheme destroys O$_3$ most effectively from 20 to 60\,km altitude (see green solid curve in Fig.~\ref{fig:chemspe}, bottom left), while the HOx scheme is most effective from 10 to 40\,km (see solid light solid blue curve in Fig.~\ref{fig:chemspe}, bottom left). 
 The catalytic NOx-cycle significantly outweighs the HOx cycle. While both separately remove O$_3$, when combined (4.0, solid yellow line, Fig.~\ref{fig:chemspe}), HOx and NOx production schemes cause an increased O$_3$ profile compared with the dominant NOx cycle of scenario 4.1. This increase in O$_3$ arises from the storage of both NOx and HOx O$_3$-destroying species in the  unreactive reservoir species HNO$_3$, formed mainly via
\begin{align}
\text{NO$_2$ + OH + M } & \rightarrow \text{ HNO$_3$ + M}\,.
\end{align}
Contrary to the O$_3$ destruction result, NOx and HOx production schemes reinforce each other with respect to destroying CH$_4$. This is likely due to both schemes producing OH, either via direct H$_2$O destruction (HOx scheme), through the NOx scheme via, e.g. 
\begin{align}
\text{NO + HO$_2$ } & \rightarrow \text{OH + NO$_2$} \, ,
\end{align}
which is ten times faster for scenario 4 compared to \citet{grenfell2012response} (scenario 6) due to increased HO$_2$ availability provided by the new scheme.

Figure \ref{fig:chemspe} also shows that the additional CR-produced NO of scenarios 4.0, 4.1, and 4.3 causes an increased  O$_3$ concentration in the troposphere with regard to scenario 1. 
The increased availability of NOx is known to favour the Smog mechanism \citep[see e.g.][]{haagen1952chemistry,grenfell2006potential}, which is responsible for most tropospherically produced O$_3$ on Earth.

\subsection{Planetary spectra} \label{sec:plaspec}

\subsubsection{Emission spectra}

We calculate theoretical emission and transmission spectra analogous to \citet{rauer2011potential}. For exoplanets, planetary emission spectra are best measured close to the secondary transit, where the insolated side of the planet is directed towards the observer. Since it is difficult to separately resolve  transiting exoplanets from their host star, therefore the measured flux $F_1$ is
\begin{align}
F_1(\lambda) = F_p(\lambda) + F_s(\lambda) \text{ ,}
\end{align}
where $F_p$ is the planetary emission flux and $F_s$ the stellar flux depending on the wavelength $\lambda$.
To obtain $F_p$, another flux has to be measured with $F_2(\lambda)=F_s(\lambda)$, which is realised when the star occults the planet during secondary transit \citep{rauer2011potential}. The planetary emission flux is then calculated via 
\begin{align}
F_p(\lambda) = F_1(\lambda) - F_2(\lambda) \text{ ,}
\end{align}
where $F_1$ and $F_2$ need to have the same integration times to remain comparable.

The contrast $C_{p/s}$ between the planetary and stellar spectrum is calculated by
\begin{align}
C_{p/s}(\lambda) = \frac{F_p(\lambda)}{F_s(\lambda)} = \frac{R_p^2}{R_s^2} \frac{I_p}{I_s}
\end{align}
with the planetary radius $R_p$, the stellar radius $R_s$, and the spectral flux densities $I_p$ and $I_s$ of the planet and the star, respectively 
\citep{rauer2011potential}. 

\begin{figure}[]
\includegraphics[width=\columnwidth]{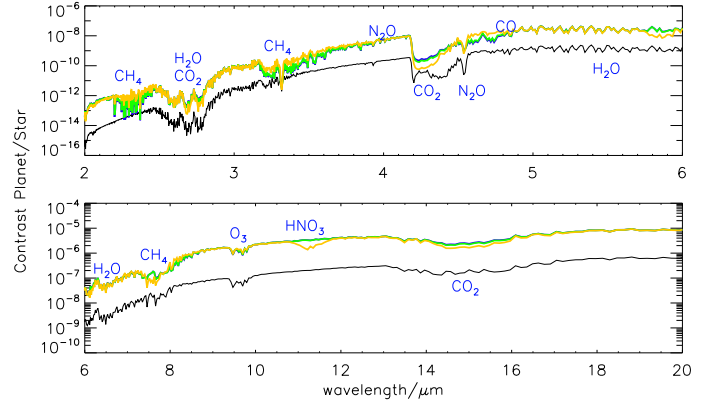}
\caption{Planetary emission contrast spectra (with resolution $R=2000$) for scenarios 0 (Earth GCR: black), 1 (M-dwarf w/o CR: blue), 3 (active M-dwarf: green), and 4 (flaring M-dwarf: yellow). The blue curve is concealed behind the green curve for most wavelengths.}
\label{fig:em}
\end{figure}

Figure \ref{fig:em} shows the planet-to-star contrast $C_{p/s}(\lambda)$ due to emission. We have plotted scenarios 0, 1, 3, and 4, although only scenario 4 shows significant changes with respect to scenario 1. As expected from the chemistry section, the CH$_4$ signals at  2.3 and 3.4\,$\mu$m decrease slightly for the flaring scenario (scenario 4), yet they still remain visible because even though 80\% of all atmospheric CH$_4$ is destroyed in the simulated M-star planet for scenario 4, there  still remains 50 times more CH$_4$ than on Earth. Additionally, the O$_3$ band at 9.6\,$\mu$m gets weaker, but is still recognizable for the flaring case (scenario 4). An HNO$_3$ signal becomes visible at around 11.2\,$\mu$m, consistent with the increased HNO$_3$ concentration. Spectral signals of CO$_2$ increase in strength. This is a temperature effect, as a colder stratosphere provides less intensity in the 15 $\mu$m CO$_2$ emission band \citep[see e.g.][]{takashi1959}. 

\subsubsection{Transmission spectra}

Transmission spectra are obtained through the measurements during and after the primary transit of the planet, where the planet occults a part of the star. In this section theoretical transmission spectra $T_i(\lambda)$ are  calculated. Therein we assume that beams cross the atmosphere at equidistant tangent heights corresponding to the atmosphere model layers. 
The total transmission $T$ is then determined by the sum of individual transmission beams $T_i$, i.e. 
\begin{align}
T=\frac{1}{H} \sum_i T_i \Delta h_i
\end{align}
where $\Delta h_i$ is the distance between tangent heights and H is the maximum height of the modelled atmosphere. 

For the Earth around the Sun a transmission spectrum may lead to higher signal-to-noise in the visible due to its higher stellar flux.
M-stars show larger relative fluxes in the IR, which facilitates the study of the absorption bands of relevant chemical species \citep[see][]{rauer2011potential}.

\begin{figure}[]
\includegraphics[width=\columnwidth]{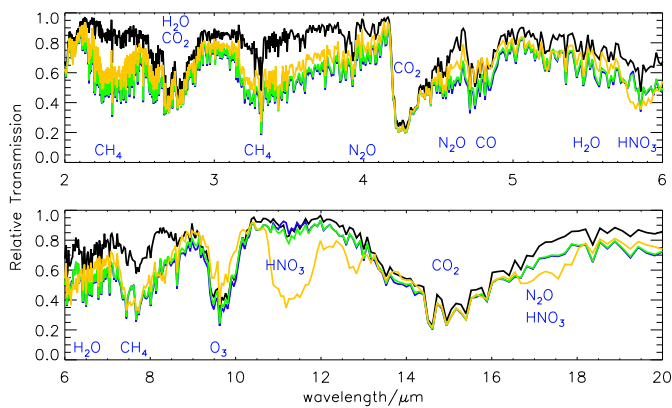}
\caption{Relative transmission spectra (with resolution $R=2000$) for scenarios 0 (Earth GCR: black), 1 (M-dwarf w/o CR: blue), 3 (active M-dwarf: green), and 4 (flaring M-dwarf: yellow). The blue curve is concealed behind the green curve for most wavelengths. }
\label{fig:trans}
\end{figure}

Figure \ref{fig:trans} shows the relative transmission spectra of the modelled scenarios. 
For the flaring case (scenario 4), the spectra change considerably, as is expected by the large change in chemical composition presented in the previous section. Spectral signals of CH$_4$ become weaker compared to the less perturbed cases (see above).  The spectral bands of N$_2$O also become slightly weaker. The ozone band at 9.6\,$\mu$m remains visible, but is weaker than in the Earth case. This is because, in contrast to \citet{grenfell2012response}, not all O$_3$ is destroyed by NOx since the HOx source we introduced provides a sink for NOx, forming reservoir species such as HNO$_3$. As a consequence, the spectral signals of HNO$_3$ become considerably stronger. This suggests that the HNO$_3$  absorption bands at 11.2\,$\mu$m may serve as a spectral marker for CRs in Earth-like atmospheres. To confirm this, further study is required regarding the range of possible conditions (UV, NOx, CR intensity) for which the HNO$_3$ signal is visible in the theoretical spectra.

\subsubsection{The very strongly flaring scenarios}  \label{sec:flare}

\begin{figure}[]
\includegraphics[width=\columnwidth]{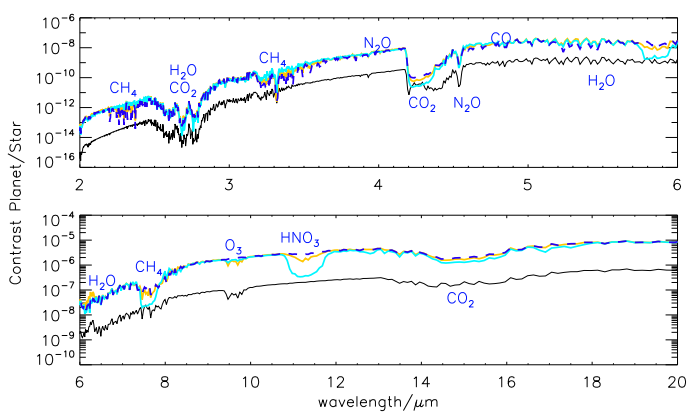}
\caption{Planet to star contrast  spectra (with resolution $R=2000$) for scenarios 0 (Earth GCR: black), 4 (flaring M-dwarf: yellow), 5 (flaring(x40) M-dwarf: light blue), and 6 (flaring M-dwarf\citep{grenfell2012response}: dark blue dashed).}
\label{fig:emtransx1}
\end{figure}

\begin{figure}[]
\includegraphics[width=\columnwidth]{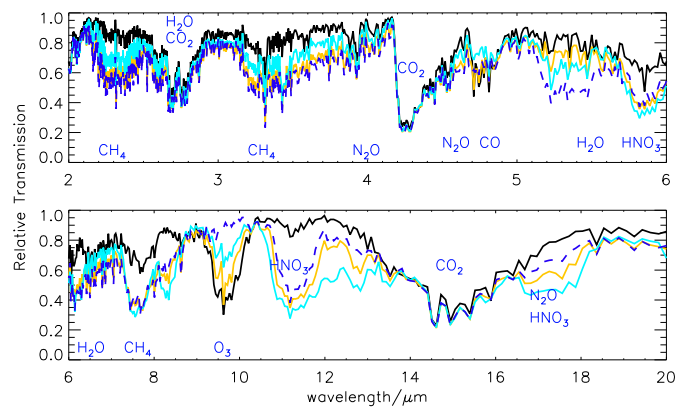}
\caption{Relative transmission spectra (with resolution $R=2000$) for scenarios 0 (Earth GCR: black), 4 (flaring M-dwarf: yellow), 5 (flaring(x40) M-dwarf: light blue), and 6 (flaring M-dwarf\citep{grenfell2012response}: dark blue dashed).}
\label{fig:emtransx2}
\end{figure}

In this section we show planetary spectra for the comparative flaring scenarios (4-6). As mentioned earlier the IPR profile produced by our updated scheme was roughly 40 times smaller than the IPR from our previous scheme \citep{grenfell2012response}. We are confident that our current updates have improved the way we model SCRs. Nevertheless, we not only wish to compare our updated CR scheme directly to our previous one, but also cover the whole spectrum of possible maximum flaring scenarios present in the literature \citep[see e.g.][]{grenfell2012response}  to see how our updated chemistry scheme behaves when fed with these high ion pair production rates. 

Figures \ref{fig:emtransx1} and \ref{fig:emtransx2} show the emission and transmission spectra, respectively, for the flaring scenarios (4-6) as well as the Earth reference case (0). As expected, scenario 6 shows very low O$_3$. However, scenario 5 even though it exhibits only 40 DU of tropospheric O$_3$, shows a clear O$_3$ signal in the relative transmission spectra, and a very slight dip in the contrast spectrum. Although somewhat weaker than the O$_3$ signal of scenario 4, this result shows that for all assumed CR intensities, O$_3$ remains visible in the theoretical planetary spectra when incorporating CR-induced N and HOx production. Also the HNO$_3$ signal becomes significantly larger for scenario 5. 

\section{Surface UV flux} \label{sec:sflux}

In the previous section, we have seen how SCRs may reduce the atmospheric ozone shield. Now we check whether this reduced ozone shield is still able to efficiently protect the planetary surface against harmful UV radiation. 

The UV radiation on the planet's surface could be harmful to developing and existing life. Especially UV-C radiation (for our model 176 - 280\,nm) and high doses of UV-B (315 - 280\,nm) can damage organic molecules and, thus, be harmful for life.

Similar to \citet{griessmeier2014}, we numerically study the full UV spectrum from $176-315$\,nm, and present integrated results for the two relevant bands UV-A and UV-B. The UV-C fluxes turn out to be negligible as a result of the efficiency of atmospheric shielding in this wavelength range.

\subsection{Surface flux calculation}

Here, we give a brief description of the way we calculate the surface UV flux.
More details can be found in \citet{griessmeier2014}.
\begin{itemize}
	\item	
		We take the wavelength-resolved stellar UV flux 
		and, with the planetary orbital distance, calculate 
		the flux incident at TOA of the M-dwarf star planet.
		From this we calculate the wavelength-integrated TOA UV fluxes  	
		$I^{\text{UV-A}}_\text{TOA}$ and $I^{\text{UV-B}}_\text{TOA}$. 
	\item	
		For this UV flux, using the numerical model described in Sect.~\ref{sec:atm}, 
		we calculate the wavelength-resolved flux of UV at the planetary surface. 
		From this, we calculate the wavelength-integrated surface UV fluxes  
		$I^{\text{UV-A}}_\text{surface}$ and $I^{\text{UV-B}}_\text{surface}$. 
	\item	
		We calculate the ratio of UV penetrating through the planetary atmosphere, averaged over 
		the corresponding UV band,
		i.e.~$R^{\text{UV-A}}=I^{\text{UV-A}}_\text{surface} / I^{\text{UV-A}}_\text{TOA}$ and 
		$R^{\text{UV-B}}=I^{\text{UV-B}}_\text{surface} / I^{\text{UV-B}}_\text{TOA}$.
		Thus, $R$  characterised the average UV shielding by the atmosphere in a particular 
		wavelength band.
	\item
		We multiply the wavelength-resolved UV surface spectra with the DNA action spectrum of 
		\citet[][Fig.~1]{cuntz2009biological}
		to calculate the effective biological UV flux $W$ at 
		the planetary surface.
		In this, the DNA action spectrum is normalised to 1 at a wavelength of 300 nm.
\end{itemize}	

We examine the UV flux for the main scenarios from Sect.~\ref{sec:crs}: scenarios 1 (M-dwarf w/o CR), 2 (quiescent M-dwarf), 3 (active M-dwarf), and 4 (flaring M-dwarf). In addition, we compare these scenarios to a scenario with Galactic (rather than stellar) cosmic rays, taking the results from \citet{griessmeier2014}. For all cases discussed in this section, we  assume unmagnetised planets, following the argumentation of \citet{griessmeier2014a} and \citet{griessmeier2014} that magnetic fields on super-Earths around
M-dwarf stars are likely to be weak and short-lived.

For each of these scenarios, we examine three different cases for the UV flux: quiescent UV (as is assumed throughout the rest of this article), short UV flare, and long UV flare:
\begin{itemize}	
\item{Quiescent UV:}
We use the TOA UV spectrum of AD Leonis, as described in Sect.~\ref{sec:atm}.

\item{Short UV flare:}
During a stellar flare, the UV flux increases by one order of 
magnitude in a typical case, 
and by at least two orders of magnitude 
in more extreme cases with a timescale of 10$^2$-10$^3$ seconds. 
We use the TOA UV spectrum of 
\citet[][Fig.~3, bold blue line, scaled for distance]{segura2010effect}, i.e.~the maximum flux at the flare peak. 
Our model does not feature time dependence. Therefore, if the timescale of the UV flare (e.g.~its duration) is short 
compared to the atmospheric reaction time, we assume that the atmosphere has not yet adjusted to the 
 increased UV flux. In this 
case, we use the atmospheric transfer function $R(\lambda)$ 
obtained in the quiescent UV case, and multiply this function with the TOA flare UV flux. 
This scenario is appropriate for an isolated flare.

\item{Long UV flare:}
The flare timescale is longer than the typical 
reaction time of the planetary atmosphere where it is assumed that the atmosphere is continuously perturbed. 
In this case, we calculate the surface UV flux using the flare TOA flux via the model of Sect.~\ref{sec:atm}. This scenario is appropriate for quasi-continuous flares.
\end{itemize}

\subsection{Surface flux results} \label{sec:sres}

\begin{table*}[]
\centering
%\begin{sidewaystable*}[p]
\begin{center} 
\small{
   \begin{tabular}{|c|c||ccc|ccc|ccc|}\hline
        %1 & column 2 & 3 & 4 & 5 \\[4pt] \hline \hline
        case \# & UV-A  		
	& &quiescent UV& & &short UV flare&& &long UV flare&	   \\   
	& & $I^{\text{UV-A}}_\text{TOA}$ & $I^{\text{UV-A}}_\text{surface}$ & $R^{\text{UV-A}}$
 	  & $I^{\text{UV-A}}_\text{TOA}$ & $I^{\text{UV-A}}_\text{surface}$ & $R^{\text{UV-A}}$
 	  & $I^{\text{UV-A}}_\text{TOA}$ & $I^{\text{UV-A}}_\text{surface}$ & $R^{\text{UV-A}}$ \\[2pt]
	& & [W/m$^2$] & [W/m$^2$] & & [W/m$^2$] & [W/m$^2$] & & [W/m$^2$] & [W/m$^2$] & \\[4pt] \hline \hline
	1 & M-dwarf w/o CR
		& 2.1 & 1.5 & 0.72
		& 82 & 57 & 0.69
		& 82 & 76 & 0.92
		\\[4pt] %\hline \hline
	$^\star$ & GCR
		& 2.1 & 1.5 & 0.72
		& 82 & 57 & 0.69
		& 82 & 76 & 0.92
		\\[4pt] %\hline \hline
	2 & SCR (quiescent M dwarf)
		& 2.1 & 1.5 & 0.72
		& 82 & 57 & 0.69
		& 82 & 76 & 0.92
		\\[4pt] %\hline \hline
	3 & SCR (active M dwarf)
		& 2.1 & 1.5 & 0.72
		& 82 & 57 & 0.69
		& 82 & 76 & 0.92
		\\[4pt] %\hline \hline
	4 & SCR (flaring M dwarf)
		& 2.1 & 1.5 & 0.72
		& 82 & 56 & 0.69
		& 82 & 79 & 0.97
		\\[4pt] %\hline \hline
\hline
   \end{tabular}
   \caption[]
   {UV-A flux at top of atmosphere ($I^{\text{UV-A}}_\text{TOA}$), at the surface ($I^{\text{UV-A}}_\text{surface}$), and 
wavelength-range averaged atmospheric transmission coefficient $R^{\text{UV-A}}=I^{\text{UV-A}}_\text{surface}/I^{\text{UV-A}}_\text{TOA}$.
Notes: $^\star$ results taken from 
\citet[][their cases CA, LF and SF, each with $\mathcal{M}=0$]{griessmeier2014}.
   }
\label{tab-uv-a}
}
\end{center} 
\end{table*}

\begin{table*}[]
\centering
%\begin{sidewaystable*}[p]
\begin{center} 
\small{
   \begin{tabular}{|c|c||ccc|ccc|ccc|}\hline
        %1 & column 2 & 3 & 4 & 5 \\[4pt] \hline \hline
        case \# & UV-B 		
	& &quiescent UV& & &short UV flare&& &long UV flare&	   \\   
	& & $I^{\text{UV-B}}_\text{TOA}$ & $I^{\text{UV-B}}_\text{surface}$ & $R^{\text{UV-B}}$
 	  & $I^{\text{UV-B}}_\text{TOA}$ & $I^{\text{UV-B}}_\text{surface}$ & $R^{\text{UV-B}}$
 	  & $I^{\text{UV-B}}_\text{TOA}$ & $I^{\text{UV-B}}_\text{surface}$ & $R^{\text{UV-B}}$ \\[2pt]
	& & [W/m$^2$] & [W/m$^2$] & & [W/m$^2$] & [W/m$^2$] & & [W/m$^2$] & [W/m$^2$] & \\[4pt] \hline \hline
	1 & M-dwarf w/o CR
		& 0.18 & 0.020 & 0.11
		& 49 & 4.8 & 0.10
		& 49 & 3.1 & 0.06
		\\[4pt] %\hline \hline
	$^\star$ & GCR
		& 0.18 & 0.023 & 0.13
		& 49 & 5.3 & 0.11
		& 49 & 3.1 & 0.06
		\\[4pt] %\hline \hline
	2 & SCR (quiescent M dwarf)
		& 0.18 & 0.021 & 0.12
		& 49 & 4.8 & 0.10
		& 49 & 3.1 & 0.06
		\\[4pt] %\hline \hline
	3 & SCR (active M dwarf)
		& 0.18 & 0.028 & 0.16
		& 49 & 4.8 & 0.10
		& 49 & 3.1 & 0.06
		\\[4pt] %\hline \hline
	4 & SCR (flaring M dwarf)
		& 0.18 & 0.05 & 0.28
		& 49 & 4.8 & 0.10
		& 49 & 11.4 & 0.23
		\\[4pt] %\hline \hline
\hline
   \end{tabular}
   \caption[]
   {As table \ref{tab-uv-a}, but for UV-B.}
\label{tab-uv-b}
}
\end{center} 
\end{table*}

\begin{table*}[]
\centering
%\begin{sidewaystable*}[p]
\begin{center} 
\small{
   \begin{tabular}{|c|c||c|c|c|}\hline
        %1 & column 2 & 3 & 4 & 5 \\[4pt] \hline \hline
        case \# & biologically weighted surface flux 
		& quiescent UV & short UV flare & long UV flare	   \\   
        & $W$  [W/m$^2$] (compared to Present Day Earth)
		& & &  
		\\[4pt] \hline \hline
	1 & M-dwarf w/o CR
		& 0.0015 (0.01 PDE$^\dagger$) & 0.48 (4 PDE$^\dagger$) & 0.15 (1.2 PDE$^\dagger$)
		\\[4pt] %\hline \hline
	$^\star$ & GCR
		& 0.0021 (0.02 PDE$^\dagger$) & 0.61 (5 PDE$^\dagger$) & 0.15 (1.2 PDE$^\dagger$)
		\\[4pt] %\hline \hline
	2 & SCR (quiescent M dwarf)
		& 0.0017 (0.01 PDE$^\dagger$) & 0.48 (0.01 PDE$^\dagger$) & 0.15 (1.2 PDE$^\dagger$)
		\\[4pt] %\hline \hline
	3 & SCR (active M dwarf)
		& 0.15 (1.2 PDE$^\dagger$) & 0.48 (0.01 PDE$^\dagger$) & 0.17 (1.3 PDE$^\dagger$)
		\\[4pt] %\hline %\hline
	4 & SCR (flaring M dwarf)
		& 0.025 (0.2 PDE$^\dagger$) & 0.48 (0.01 PDE$^\dagger$) & 4.9 (40 PDE$^\dagger$)
		\\[4pt] %\hline %\hline
\hline
   \end{tabular}
   \caption[]
   {Biologically weighted surface UV-flux, in weighted W/m$^2$ and relative to present day Earth.
Notes: $^\star$ results taken from \citet[][their cases CA, LF and SF, each with $\mathcal{M}=0$]{griessmeier2014}. 
$^\dagger$ PDE=Present Day Earth ($W$=0.126 W/m$^{2}$, see \citet{griessmeier2014})
   }
\label{tab-w}
}
\end{center} 
\end{table*}

Tables \ref{tab-uv-a}, \ref{tab-uv-b}, and \ref{tab-w} show the UV-A flux (315-400\,nm), the UV-B flux (280-315\,nm), and the biologically weighted UV-flux $W$
for scenarios 1 through 4 plus, for comparison, the Galactic cosmic ray scenario from \citet{griessmeier2014}. For each scenario, we also compare the three different UV cases discussed above.

As seen in Table \ref{tab-uv-a}, the atmospheric transmission coefficient $R$ for UV-A is virtually independent of the cosmic ray environment. It does, however, vary with the ambient UV conditions. During a long UV-flare, the surface receives up to 50 times more UV-A radiation than under quiescent conditions.

Table \ref{tab-uv-b} shows that the UV-B flux is more sensitive to the destruction of the ozone layer by CRs. The presence of a strong cosmic ray environment can increase the atmospheric transmission coefficient for UV-B (and the surface UV-B flux) by a factor of almost 4. Still, the variability of the surface UV flux is dominated by the UV activity of the host star, which can change the surface UV-B flux by over two orders of magnitude.

The biological radiation damage created by UV-A radiation is several orders of magnitude weaker than the damage caused by UV-B radiation \citep[e.g.][]{cuntz2009biological}. 
Thus, it is not surprising to see that the biologically weighted UV flux $W$ qualitatively shows a similar behaviour as the UV-B flux. Quantitatively, the response of $W$ is stronger than that of the average UV-B flux. This results from the fact that $W$ is is sensitive mostly to UV-B flux at 300\,nm, which is more variable than the band-averaged UV-B flux.
Table \ref{tab-w} shows that under high cosmic ray fluxes (i.e.~during a strong stellar particle flare), the planetary ozone shield is reduced strongly enough to increase the weighted surface UV flux $W$ by a factor of 20. During a long stellar UV-flare, $W$ increased by an additional factor of 100. 
When compared to Earth, the value of $W$ remains below the terrestrial value for almost all cases.
However, when the high UV flux of a long stellar flare coincides with a stellar particle eruption (which reduces the atmospheric UV shield), $W$ reaches values considerably higher than on present day Earth (40 times the present-day terrestrial value).

Our results compare well to previous work, and extend them by including the increase in stellar UV flux. 
\citet{grenfell2012response} consider the effect of the particle flux on the atmospheric ozone layer, but do not take the increase in stellar UV during a flare into account. \citet{grenfell2012response} find that the surface flux of UV-B (their Table 2) is increased by a factor of $\sim30$ in the flaring case. Even with this, the surface UV-B flux is considerably lower that the terrestrial value, which is qualitatively consistent with our quiescent UV case. 
In the current work (using the new modelling scheme), less ozone is destroyed by SCRs, so that the surface is better protected against UV-B. However, in the new cases of stellar UV flares (both short and long), this is more than compensated by the increase in stellar UV flux, leading to an increase in surface UV-B by several orders of magnitude.

Our results are also compatible with those of \citet{segura2010effect}. 
They considered both the flare's increased UV radiation and a proton event at the peak of the flare.
Their flare peak is comparable to our short UV-flare case for the flaring M-dwarf scenario (see their Table 2), and the UV-B surface flux values are indeed comparable. Our work adds the long UV-flare case, which has a higher potential impact on habitability, and the analysis of the biologically weighted surface UV flux.

\subsection{Surface flux: Discussion}

Our main findings are as follows. 
We find that CRs leave the UV-A transmission coefficient and flux virtually unchanged. For UV-B, GCRs modify the transmission rate and surface flux by less than 20\%, whereas SCRs change $R$ by more than a factor of two. For the GCR scenario, short flares are more harmful than long flares. For the SCR scenarios, long flares (where $R$ has time to adjust to the modified particle flux) are more harmful than short flares.

Regarding the biologically weighted UV surface flux $W$, GCRs can change $W$ by up to 40\%, while SCRs can change $W$ by over one order of magnitude. In all cases, $W$ is dominated by the contribution of UV-B. In the case of Earth (without any CRs), $W=0.125$ W/m$^2$ \citep{griessmeier2014}. Except in the long UV flare case, the corresponding value for an M-dwarf, even when subject to strong cosmic ray fluxes, is lower. With  $W\le 4.9$ W/m$^2$ (Table \ref{tab-w}), the SCR-induced surface UV radiation during a long stellar UV flare may indeed potentially be harmful for certain types of cells (e.g. human skin or eye cells), but is not strong enough to prevent life. 
For example, \textit{Deinoccocus radiodurans} is able to withstand a flux of $>40$ W/m$^{2}$ without significant damage, and life on Earth may have arisen during times when the biologically weighted UV flux was even higher \citep[$>96$ W/m$^{2}$, see][]{cockell1999carbon}.

\section{Discussion} \label{sec:disc}

In the context of the photochemistry of Earth-like atmospheres the unconstrained parameter range is large. Key parameters include atmospheric pressure \citep[e.g.][]{vladilo2013habitable}, composition \citep[e.g.][]{segura2007abiotic}, incoming stellar spectrum \citep[e.g.][]{segura2003ozone}, position in the habitable zone \citep[e.g.][]{grenfell2007response}, incoming cosmic ray fluxes \citep[e.g.][]{griessmeier2014a}. 
Our model results can be compared best to the work of \citet{grenfell2012response} as we use the same radiative-convective/photochemical model, but  with an improved CR scheme (see Sect.~\ref{sec:imp}). A comparison shows that 
the improved scheme conserves spectrally visible amounts of O$_3$ in the studied atmospheres and catalytic O$_3$ destruction is partly mitigated, helped also by an NOx-induced O$_3$-producing smog mechanism in the troposphere. Another application of a photochemical model to exoplanets includes a study by \citet{segura2005biosignatures}, who studied the effect of an M-dwarf spectrum on an Earth-like atmosphere. Their placement of the planet within the HZ differs from the current work in that their total stellar flux was normalised so that the surface temperature equals 288\;K.  \citet{segura2005biosignatures} find large concentrations of CH$_4$ (Earth-like biogenic flux, but decreased photolysis of CH$_4$), causing a strong CH$_4$ greenhouse effect. Adding CR-induced OH production to our scheme provides a further sink for the high CH$_4$ levels. This causes a reduction of CH$_4$ concentrations by a factor of five in the flaring case (4) as well as a reduction in the surface temperature by 7\;K.
\citet{segura2010effect} have studied the effect of an M-dwarf stellar flare. As mentioned in section \ref{sec:sres} the UV-B values of our short UV-flare case for the flaring M-dwarf scenario and their flare peak are comparable.
Further photochemical studies regarding planets orbiting M-dwarf planets include \citet{tian2014}, who use an early-Earth model with variable O$_2$ and high CO$_2$ concentrations, whereas we assume an Earth-like setup with constant N$_2$, O$_2$, CO$_2$ and biogenic fluxes. While \citet{tian2014} study the abundance of O$_2$  under abiotic conditions for different CO$_2$ mixing ratios, we study the effect of (in this case) CRs on bioindicator species and trace gases under biotic and Earth-like conditions. Additionally, \citet{tian2014} use the stellar spectrum of an M-dwarf with low UV flux due to recent measurements of quiescent M-dwarfs \citep{france2012time,france2013ultraviolet}, whereas we use the active M-dwarf spectrum of AD Leo to provide comparison to previous works and because the current work focuses on the influence of the changes to the CR scheme. The impact of different M-dwarf spectra to our photochemical model was studied by \citet{grenfell2013potential,grenfell2014sensitivity}. \citet{grenfell2013potential} presented a detailed analysis of atmospheric pathways that form and destroy, e.g. O$_3$, for an Earth-like planet orbiting in the HZ of different M-dwarf stars.  \citet{grenfell2014sensitivity} studied the effect of varying the (uncertain) incoming UV flux for Earth-like planets orbiting in the HZ of M-dwarfs. They found that while increases in UV$_{\text{\textasciitilde}200-300\text{nm}}$ favoured O$_3$ loss via photolysis, increases in UV$_{< 200 \text{nm}}$ favoured O$_3$ production by stimulating O$_2$ photolysis.

Several caveats of our model require discussion. 
The parametrisations used in the chemistry improvements were set up for the Earth atmosphere only, and are hence valid when assuming a nitrogen-oxygen dominated atmosphere with Earth's mixing ratios. Calculating other atmospheric compositions would require input from an ion chemistry model to infer IPR as well as production coefficients.

The changes to our air-shower scheme have caused the IPR of our constant flaring scenario to decrease by a factor of 40. 
However, these changes have made our model more physically realistic, as primary particles of different energies now produce decidedly different amounts of secondary electrons at different altitudes, which was only partially true for the old scheme. In addition, we have shown that the changes to our chemical scheme with regard to CRs lead to less photochemical O$_3$ loss, so that this species is then visible in the theoretical spectra, even for an IPR increased by a factor of 40 (see Sect.~\ref{sec:flare}).

It is challenging to apply our CR air-shower scheme  to the relatively low energies of primary proton fluxes originating from the planet's host star (SCRs). 
We need better information on ion pair production rates for low-energy cosmic ray particles are, to better constrain the interpolation from higher to lower energies. 

Our atmospheric model operates in steady state. Consequently, time-resolved flaring scenarios are not possible. Nevertheless, for strongly flaring stars such as AD Leo, we generally assume rapid flaring compared to photochemical response timescale suggesting a (quasi-)stationarity. Since such a constant flare output would in reality still vary largely in intensity, our constant flaring scenario is therefore to be considered as an upper boundary of the possible effect of SCRs on the atmospheric chemistry of Earth-like planets in the HZ of flaring M-dwarfs.

\section{Summary and conclusion} \label{sec:summ}

We improved upon the CR scheme of \citet{grenfell2012response} by updating the CR parametrisation parameters, as well as adding CR-induced production of N and HOx to the existing NO production scheme via ion pair production parametrisations obtained from Earth measurements by  \citet{jackman1980production} and \citet{solomon1981effect}. With these improvements, we have shown that the atmospheric effects of high CR fluxes do not completely destroy O$_3$ reservoirs as previously calculated \citep[see][]{grenfell2012response}, but leave enough O$_3$ such that it remains visible in theoretical planetary spectra. The atmospheric accumulation of CR-produced NOx and HOx leads to a very large HNO$_3$ reservoir, whose signal is very clear in the theoretical spectra.

Our main results are that SCR-induced NOx and HOx production for Earth-like planets orbiting in the HZ of M-dwarfs removes some of the produced NO via storage as HNO$_3$, thereby weakening the catalytic destruction of O$_3$. We find that HNO$_3$ may be a potentially detectable indicator of high incident stellar particle flux, yet further investigation is needed. Additionally, the new CR chemistry scheme helps destroy significant amounts of CH$_4$  due to increased OH production. Both O$_3$ and CH$_4$ remain visible in the modelled planetary spectra even when assuming a constant flaring of the host star. 

If a stellar cosmic ray particle event coincides with a stellar UV flare, the surface UV flux may strongly exceed that of present-day Earth. This flux level exceeds that found previously \citep[][]{griessmeier2014} for the case of GCRs. Still, even if the SCR-induced surface UV radiation 
may indeed be potentially harmful for certain types of cells, 
it is not strong enough to prevent life.

\bibliographystyle{aa}
\bibliography{paper2}

\end{document}